\documentclass[a4paper,11pt]{article}

\usepackage{jheppub} % for details on the use of the package, please
\usepackage{longtable}
\usepackage{latexsym}
\usepackage{hyperref}
\usepackage{makeidx}
\usepackage{amssymb}
\usepackage{amsmath}
\usepackage[all,cmtip]{xy}
\xyoption{2cell}
\UseAllTwocells

\usepackage{rotating}

\newcommand{\cR}{{\cal R}}
\newcommand{\I}{{\cal I}}
\newcommand{\half}{\frac{1}{2}}

\usepackage{tikz}
\usetikzlibrary{cd}
\usepackage{pgfplots}
\usetikzlibrary{decorations.markings}
\usetikzlibrary{decorations.pathmorphing}
\hypersetup{
 colorlinks,
 citecolor=blue,
 filecolor=blue,
 linkcolor=blue,
 urlcolor=blue
}
\definecolor{shadecolor}{rgb}{0.8,0.9,0.9}

\newcommand{\K}{{\cal K}}
\newcommand{\V}{{\cal V}}
\newcommand{\F}{{\cal F}}
\newcommand{\A}{{\cal A}}
\newcommand{\N}{{\cal N}}
\newcommand{\Ham}{{\cal H}}

\usepackage{cancel}

\title{\boldmath Type-II Calabi-Yau compactifications, T-duality and special geometry in general 
spacetime signature}

\author{M. M\'edevielle}
\author{T. Mohaupt}
\author{and G. Pope}
\affiliation{Department of Mathematical Sciences \\
University of Liverpool\\
Peach Street, Liverpool L69 7ZL, UK}

\emailAdd{maxime.medevielle@liverpool.ac.uk}
\emailAdd{Thomas.Mohaupt@liv.ac.uk}
\emailAdd{giacomo@liverpool.ac.uk}

\abstract{
We obtain the bosonic Lagrangians of vector and hypermultiplets
coupled to four-dimensional $\mathcal{N}=2$ supergravity in 
signatures $(0,4)$, $(1,3)$ and (2,2) by compactification of 
type-II string theories in signatures (0,10), (1,9) and (2,8) 
on a Calabi-Yau threefold. 
Depending on the signature and 
the distinctions between type-IIA/IIA$^*$/IIB/IIB$^*$/IIB' the resulting
scalar geometries are special K\"ahler or special para-K\"ahler
for vector multiplets and quaternion-K\"ahler or para-quaternion
K\"ahler for hypermultiplets. By spacelike and timelike reductions
we obtain three-dimensional $\mathcal{N}=4$ supergravity theories coupled
to two sets of hypermultiplets. We determine the c-maps relating
vector to hypermultiplets, and show how the four-dimensional theories
are related by spacelike, timelike and mixed, signature-changing 
T-dualities. }

\begin{document}

\maketitle
\flushbottom

\section{Introduction}

String theory is a web of perturbatively defined theories 
which are related to each other by various dualities. 
In particular, ten-dimensional type-II string theories, which 
have the maximal amount of supersymmetry, are related 
to each other by T-duality and S-duality. If one includes
timelike T-duality, then besides the familiar type-IIA and 
type-IIB theories there exist two further theories in Lorentz signature,
type-IIA$^*$ and type-IIB$^*$, and there also exist 
further type-II theories in all possible ten-dimensional 
spacetime signatures \cite{Hull:1998vg,Hull:1998ym,Hull:1998fh}. 
The formal properties of these theories as well as their
potential applications in model building and cosmology have
been investigated further in \cite{Dijkgraaf:2016lym,Blumenhagen:2020xpq}.
Exotic type-II theories have unusual features and their ultimate
role in string theory remains to be understood. 
From the point of view
of symmetries and string geometry, it is natural to include them. 
Timelike dimensional reduction is a valid solution-generating technique, 
and timelike T-dualities exist whenever one can find an alternative
dimensional up-lift. Symmetries which become manifest in 
dimensional reduction give information about the hidden 
symmetries of the full theory \cite{deWit:1985iy}. Including 
time in the reduction as a strategy for uncovering the 
full symmetry structure underlying string theory has been
advocated in \cite{Moore:1993zc}. 
In the frameworks of doubled and exceptional geometry and field theory,
type-II$^*$ theories seem to be on the same footing as the conventional
ones \cite{Hohm:2011dv,Hohm:2019bba}. Since type-II and type-II$^*$ 
theories have the same Euclideanized version \cite{Hull:1998vg}, it is natural 
to think of them as resulting from the same underlying Euclidean partition function. 

When combining timelike T-duality with S-duality, string symmetries also
lead to backgrounds with non-Lorentzian signatures. While their interpretation
is challenging, they cannot be discarded ad hoc, since 
string theory is believed to be a single theory with all of its consistent 
backgrounds connected by physical processes.
The relevant question is therefore whether vacua with exotic signature
can be generated, and evidence for this has been presented in \cite{Dijkgraaf:2016lym}. 
It has also been argued that string theories in exotic signature
can be defined holographically as duals of gauge theories based on Lie 
supergroups \cite{Dijkgraaf:2016lym}. We also note that the network of 
type-II string theories and of the related eleven-dimensional M-theories
realizes all possible ten- and eleven-dimensional supersymmetry 
algebras with 32 real supercharges \cite{Gall:2021tiu}, so that when 
allowing exotic signatures in string theory, all maximally symmetric 
supergravities in all signatures can be realized as limits. Finally,
the inclusion of non-Lorentzian signatures is natural from the point 
of view of the Euclidean approach to quantum 
gravity, since complex saddle points contribute to the functional
integral. Recently, the role of complex spacetime metrics in 
quantum field theory and quantum gravity has been emphasised 
in \cite{Kontsevich:2021dmb,Witten:2021nzp,Lehners:2021mah}. 
As a natural extension, one can complexify
all fields, which would imply to
consider all type-II theories as part of a single complex configuration 
space. We remark that complex saddle points can contribute 
to Euclidean path integrals for scalar fields, and that there are examples
where actions with inverted kinetic terms can be viewed as
arising from manipulating integration contours in complexified
field space, see \cite{Mohaupt:2010du} for an elementary example.

Calabi-Yau compactifications of Lorentz signature type-IIA/B string theories
give rise to four-dimensional $\mathcal{N}=2$ supergravity theories with
vector and hypermultiplets \cite{Ferrara:1988ff,Bodner:1990zm,Bohm:1999uk}. 
This is a much studied class of theories which 
while not phenomenologically realistic, has rich and complex dynamics,
since the scalar geometry is not rigid but depends on functions of the 
scalar fields.
$\mathcal{N}=2$ supersymmetry still severely restricts the quantum and
stringy corrections that these functions can receive, so that one 
often can find exact non-perturbative results. Applications range from 
the study of field theories, to black holes and their entropy, and to 
the AdS/CFT correspondence. It is therefore interesting to extend these
studies to the Calabi-Yau compactifications of exotic type-II theories. 

The Calabi-Yau compactification 
of the type-IIA theory in Euclidean signature $(0,10)$ has been 
worked out in \cite{Sabra:2015tsa}. Moreover, the vector multiplet
sectors of five- and four-dimensional supergravity in arbitrary
signature have been found in \cite{Sabra:2017xvx} through 
an analysis of Killing spinor equations combined with the
reductions of eleven-dimensional supergravity theories 
in signatures (1,10), (2,9) and (5,6) on 
Calabi-Yau threefolds, followed by the reduction to four 
dimensions on spacelike and timelike circles. 

In this paper we will obtain the bosonic actions for four-dimensional $\mathcal{N}=2$ 
supergravity coupled to vector and hypermultiplets
in signatures  $(0,4), (1,3)$ and $(2,2)$  by compactification of type-II string theories
in signatures $(0,10), (1,9), (2,8)$ on Calabi-Yaur threefolds. 
Carrying out these reductions is straightforward 
since for type-IIA 
one can adapt the computations of \cite{Bodner:1990zm, Sabra:2015tsa},
while the corresponding results for type-IIB are
fixed by mirror symmetry. Therefore, our main focus is the interpretation 
of the relative sign flips between terms in the resulting four-dimensional
Lagrangians in terms of the special geometry of their vector and 
hypermultiplet manifolds. These geometries vary between signatures and
between type-II and type-II$^*$. There is an intimate 
relation between these signs and the variation of the R-symmetry 
groups of the underlying supersymmetry algebras between signatures, and
between standard and twisted (type-*) supersymmetry algebras.

As is well known, in signature $(1,3)$ the scalar geometry of 
vector multiplets coupled to supergravity is special K\"ahler, 
while the geometry of hypermultiplets is quaternionic-K\"ahler, 
see \cite{Andrianopoli:1996cm,Freedman:2012zz,LopesCardoso:2019mlj,Lauria:2020rhc} for review. 
In Euclidean signature the scalar geometry of vector multiplets
becomes special para-K\"ahler, which is reflected by a change of
the abelian factor of the R-symmetry group from $U(1)$ to $SO(1,1)$ 
\cite{Cortes:2003zd,Cortes:2009cs}. In three Euclidean dimensions
the geometry of hypermultiplets is para-quaternionic K\"ahler 
\cite{Cortes:2015wca}. In \cite{Cortes:2019mfa} it was observed 
that there exists a twisted version of the $\mathcal{N}=2$
supersymmetry algebra, which has a non-compact R-symmetry group.
The vector multiplet Lagrangian differs from the standard one by a sign flip
of some kinetic terms, which is analogous to the difference between 
type-II and type-II$^*$ theories. The same type of sign flip was 
observed in \cite{Sabra:2017xvx}, when reducing five-dimensional 
vector multiplets coupled to supergravity from signature $(2,3)$ to signature
$(1,3)$. One should therefore expect that
$\mathcal{N}=2$ theories which realize the twisted $\mathcal{N}=2$
algebra can be obtained as Calabi-Yau compactifications of type-II$^*$ 
theories. In this paper we will verify this explicitly, as part of 
obtaining a complete list of scalar 
geometries for four-dimensional $\mathcal{N}=2$ supergravity with 
vector and hypermultiplets
for all signatures
through the dimensional reduction of type-II theories on Calabi-Yau
threefolds.

In addition, we will perform all possible spacelike and timelike 
dimensional reductions from four to three dimensions. After
reduction, vector multiplets can be dualized into hypermultiplets, 
so that one obtains a scalar manifold which is the product 
of two hypermultiplet manifolds. The map relating a vector multiplet manifold to
a hypermultiplet by reduction is know as the c-map 
\cite{Cecotti:1988qn,Ferrara:1989ik}. By starting in arbitrary 
signature and including timelike reductions, one obtains
variants of the c-map, which we describe for all possible cases.
Whenever a dimensional
reduction can be combined with a different dimensional lifting
(equivalently, whenever the same three-dimensional theory 
can be obtained from two different four-dimensional theories
by reduction), this realizes a T-duality between the underlying 
type-II string theories. We map out the complete network of
spacelike, timelike and mixed T-dualities, where mixed T-dualities
combine spacelike reduction/lifting with timelike lifting/reduction 
and thus change the four-dimensional signature.

We briefly mention further motivations and future application of our
work. One is the study of solutions to four-dimensional $\mathcal{N}=2$ theories with 
twisted supersymmetry and with non-Lorentzian signature, as well
as their dimensional uplifts to ten and eleven dimensions. 
In particular, according to \cite{Gutowski:2019iyo,Gutowski:2020fzb},
there is a correspondence
between the planar cosmological solutions in $\mathcal{N}=2$
vector multiplet theories that can be embedded into type-II string theory,
and planar black hole solutions in vector multiplet theories realizing
the twisted $\mathcal{N}=2$ supersymmetry algebra, which, as we
show in this paper, can be embedded into type-II$^*$. Both 
solutions can be related to the same four-dimensional Euclidean partition function, which 
explains that their Killing horizons satisfy the same thermodynamic
relations \cite{Gutowski:2020fzb}. This is consistent with type-II and type-II$^*$
having the same Euclideanized form \cite{Hull:1998vg}. 
For other work on solutions of exotic $\mathcal{N}=2$ theories
see \cite{Klemm:2015mga,Sabra:2015vca,Gutowski:2019hnl,Sabra:2020gio,Sabra:2021omz,Sabra:2021ugi}.

Another potential application is topological string theory.  Standard Type-II
Calabi-Yau  compactifications allow two topological twists, which define two topological 
worldsheet theories, the A-model and the B-model, which are sensitive to the K\"ahler
and complex structure moduli respectively. Since Calabi-Yau compactifications of 
exotic type-II theories work analogously to standard type-II theories, and given that we
will show that the geometry of the resulting moduli spaces can be determined in
the supergravity approximation, we expect that a
world-sheet perspective for these compactifications can be developed too. 
Topological string theories encode
a subsector of the full string theories, and may also be related to a `topological
phase' of string theory, where more of its symmetries become manifest \cite{Witten:1988sy}. 
We remark that in such a topological phase, the expectation value of the spacetime metric
is zero, which makes it natural that phases with non-Lorentzian signature coexist
in the theory with conventional Lorentzian phases.

The outline of this paper is as follows. We start from the classification 
of four-dimensional $\mathcal{N}=2$ and three-dimensional 
$\mathcal{N}=4$ supersymmetry algebras and explain how most of the 
qualitative features of the scalar geometries of vector and hypermultiplets
as well as their mutual relations
by T-dualities can already be predicted by inspection of the R-symmetry
groups. We present the bosonic vector and hypermultiplet Lagrangians, 
and explain the effects of changing the supersymmetry algebra on the 
scalar geometries. This includes a brief review of special para-K\"ahler 
and para-quaternion-K\"ahler geometries, which replace the familar
special K\"ahler and quaternion-K\"ahler geometries for certain signatures.
We perform all possible spacelike and timelike reductions
from signatures (0,4), (1,3) and (2,2) to signatures (0,3) and (1,2),
and show that the six resulting c-maps which map vector multiplet manifolds to
hypermultiplet manifolds fall into three distinct classes, depending on whether
the resulting hypermultiplet manifold is quaternionic-K\"ahler, 
para-quaternionic-K\"ahler with a special K\"ahler base or 
para-quaternionic-K\"ahler with a special para-K\"ahler base.

Then we review type-II string theories in ten dimensions and catalogue
the relative sign flips of their kinetic terms. Next, we explain how
these sign flips affect compactifications on Calabi-Yau threefolds,
and obtain the corresponding sign flips of the resulting four-dimensional
vector and hypermultiplet Lagrangians. While in the main part of the 
paper we just trace the kinetic terms, we provide a full derivation in 
the appendix. Here we use that the reduction of all individual terms
is available from the work of \cite{Bodner:1990zm} on the reduction 
of type-IIA with signature (1,9) and of \cite{Sabra:2015tsa} on the
reduction of type-IIA with signature (0,10). 
Combining the results from dimensional reduction  with the
previous results on c-maps we obtain six types of T-dualities by 
identifying all possible combinations of reductions from four to three
with `oxidations' from three to four dimensions.
These T-dualities organise into two orbits,
one which relates type-IIA/IIB/IIA$^*$/IIB$^*$ through `pure'
-- that is spatial or timelike T-dualities, the other which relates
type-IIA$_{(0,10)}$/IIB'$_{(0,9)}$/IIA$_{(2,8)}$ through `mixed',
signature changing T-dualities. This separation coincides with 
the one between worldsheet theories with Lorentzian and with
Euclidean signature \cite{Hull:1998ym}.
Both orbits could only be
related through the S-duality between type-IIB$^*$ and type-IIB',
which is not expected to be valid for generic $\mathcal{N}=2$
compactifications, though it may be realized for non-generic
`$\mathcal{N}=4$-like' compactifications. 

\section{Vector and hypermultiplets in four and three dimensions}

\subsection{Supersymmetry algebras in four and three dimensions}

Four-dimensional $\mathcal{N}=2$ supersymmetry algebras, that is four-dimensional 
supersymmetry algebras with 
eight real supercharges,\footnote{In Euclidean signature this is the smallest supersymmetry algebra. Our convention is to count supersymmetries in multiples of Majorana spinors, irrespective of whether
Majorana spinor spinors exist is the particular signature. This convention is natural if one considers
supersymmetry algebras in different signatures at the same time. } 
have been classified for arbitrary signature in \cite{Cortes:2019mfa}.
They are completely characterized by their R-symmetry groups, which we list in Table
\ref{Tab:4d_susy}. While the $\mathcal{N}=2$ algebra is unique in Euclidean signature
$(0,4)$ and in neutral signature (2,2), there are two non-isomorphic algebras in 
Lorentz signature (1,3).\footnote{We use the mostly plus convention, so (1,3) means
that the metric has 3 positive eigenvalues and 1 negative eigenvalue.} Besides the standard $\mathcal{N}=2$ algebra with compact
R-symmetry group $U(2)$ there exists a second algebra with non-compact R-symmetry 
$U(1,1)$, which we will refer to as the twisted $\mathcal{N}=2$ algebra. The change of the R-symmetry 
group reflects itself in certain sign flips in the bosonic  Lagrangian \cite{Cortes:2019mfa}, which
are similar to those which distinguish type-II and type-II$^*$ string theories \cite{Hull:1998vg}. 
We will see  later that theories realizing the twisted $\mathcal{N}=2$
algebra are obtained by the compactification of type-II$^*$ string theories on Calabi-Yau
threefolds. The uniqueness of the supersymmetry algebras in Euclidean and neutral
signature reflects the uniqueness of type-II string theories in signatures (0,10) and (2,8), from 
which such theories can again be obtained as Calabi-Yau compactifications.

\begin{table}
\begin{center}
\begin{tabular}{| l | l | l | l |} \hline
Signature & R-symmetry & VM geometry & HM geometry \\ \hline
$(0,4)$ & $U(2)^* \cong SO(1,1) \times SU(2)$ & SPK & QK \\
$(1,3)$ & $U(2) \cong U(1) \times SU(2)$ & SK$_+$ & QK \\
 & $U(1,1)\cong U(1) \times SU(1,1)$ & SK$_-$ & PQK \\ 
$(2,2)$ & $GL(2,\mathbb{R} )\cong SO(1,1) \times SL^\pm(2,\mathbb{R})$ & SPK & PQK \\
\hline
\end{tabular}
\end{center}
\caption{Four-dimensional $\mathcal{N}=2$ supersymmetry algebras, their
R-symmetry groups and their scalar geometries. We use the acronyms SK = special 
K\"ahler, SPK = special para-K\"ahler, QK = quaternionic K\"ahler and PQK = 
para-quaternionic K\"ahler. See Section \ref{VM} for further explanations.
\label{Tab:4d_susy}}
\end{table}

Three-dimensional $\mathcal{N}=4$ supersymmetry algebras have been classified, for arbitrary signature 
in \cite{Gall:2021tiu}, and are again characterized uniquely by their R-symmetry groups, 
see Table \ref{Tab:3d_susy}.
The embeddings $U^*(2) \subset SO^*(4)$, $U(2) \subset O(4)$, 
$U(1,1) \subset O(2,2)$ and $GL(2,\mathbb{R}) \subset O(2,2)$ 
indicate how these algebras are related to four-dimensional 
$\mathcal{N}=2$ algebras by spacelike or timelike dimensional reduction, see
Table \ref{Table:4d_3d}. There is no candidate for a dimensional lift of the
algebra with R-symmetry $O(1,3)$. In the following sections we will review
vector and hypermultiplets, in particular, the geometry of their scalar manifolds,
and how this geometry is tied to the R-symmetry group.

\begin{table}
\begin{center}
\begin{tabular}{| l | l | l | l |} \hline
Signature & R-symmetry & HM$_1$  geometry & HM$_2$ geometry \\ \hline
$(0,3)$ & $SO^*(4) \cong SL(2,\mathbb{R}) \times SU(2)$ &PQK & QK \\
$(1,2)$ & $O(4)  \cong SU(2) \times SU(2)$ & QK & QK \\
 & $O(1,3)$ & $-$  &  $-$\\ 
& $ O(2,2) \cong SL(2,\mathbb{R}) \times SL(2,\mathbb{R})$ &   PQK & PQK \\ \hline
\end{tabular}
\end{center}
\caption{Three-dimensional $\mathcal{N}=4$ supersymmetry algebras, their
R-symmetry groups and their scalar geometries. See Section \ref{HM}
for further explanations.
\label{Tab:3d_susy}}
\end{table}

\begin{sidewaystable}
\begin{tabular}{| l | l | l | l | l | l| l |}  \hline
Signature & R-symmetry & Geometry & Reduction & R-symmetry & Geometry & c-map \\ \hline
(0,4) & 
$SO(1,1)\times SU(2)$ & 
SPK $\times$ QK & 
(0,4) $\rightarrow$  (0,3) & 
$SL(2,\mathbb{R}) \times SU(2)$ & PQK $\times$ QK & Euclidean c-map \\ \cline{1-3}
(1,3) & $U(1) \times SU(2)$ & SK $\times$ QK & (1,3) $\rightarrow$ (0,3) &
$SL(2,\mathbb{R}) \times SU(2)$ & PQK $\times$ QK & Temporal c-map \\
 & $U(1) \times SU(1,1)$ & SK $\times$ QK & (1,3) $\rightarrow$  (0,3) &
$SU(2) \times SU(1,1)$ & QK $\times$ PQK & Twisted temporal c-map \\  \cline{4-7}
& $U(1) \times SU(2)$ & SK $\times$ QK & (1,3) $\rightarrow (1,2)$ &
$SU(2) \times SU(2)$ & QK $\times$ QK & (spatial) c-map \\ 
& $U(1) \times SU(1,1)$ & SK $\times$ PQK & (1,3) $\rightarrow (1,2)$ & 
$SU(1,1) \times SU(1,1)$ & PQK $\times$ PQK & Twisted (spatial) c-map \\ \cline{1-3}
(2,2) & $SO(1,1) \times SL^\pm(2,\mathbb{R})$ & SPK $\times$ PQK & 
(2,2) $\rightarrow$ (1,2) & $SU(1,1) \times SU(1,1)$ & PQK $\times$ PQK & 
Neutral c-map \\ \hline
\end{tabular}
\caption{Dimensional reduction from four to three dimensions for all inequivalent signatures: R-symmetry groups, scalar geometries, and type of c-map. }
\label{Table:4d_3d}
\end{sidewaystable}

\subsection{Vector multiplets \label{VM}}

We start in signature (1,3) with the standard $\mathcal{N}=2$ supersymmetry 
algebra with R-symmetry
group $U(2)\cong U(1)\times SU(2)$. A vector multiplet contains a complex scalar $z$,
an $SU(2)$ doublet of spinors, and a gauge field $\mathcal{A}_\mu$. The scalar and 
gauge field are neutral under $SU(2)$.  Under the $U(1)$,
the scalars, spinors and vectors carry charges $\mp 1, \mp \frac{1}{2}, 0$
respectively. The scalar manifold is an affine special K\"ahler manifold for
rigid supersymmetry and a projective special K\"ahler manifold for
local supersymmetry. We refer to \cite{LopesCardoso:2019mlj} for a review of special geometry which
uses the same conventions and terminology as used in this paper. 
Both types of special K\"ahler geometries (SK geometries)
have in common that the K\"ahler metric $g_{\alpha \bar{\beta}}(z,\bar{z})$, 
$\alpha, \beta = 1, \ldots, n_V$ of the scalar manifold can be expressed 
in terms of a holomorphic function $\mathcal{F}(z^\alpha)$, called the prepotential. 
Special K\"ahler geometry is intimately related to the invariance of the field
equations under symplectic transformations, which generalize and contain
electric-magnetic duality transformations \cite{deWit:1984pk, deWit:1984rvr}. 
We are interested in the case
where the $n_V$ vector multiplets are coupled to $\mathcal{N}=2$ supergravity. 
The supergravity multiplet contains one further vector field $\mathcal{A}^0_\mu$. 
A simple, linear action of the symplectic group $Sp(2n_V+2,\mathbb{R})$ 
is obtained by taking certain field-dependent linear combinations $A^I_\mu$,
$I=0,1, \ldots, n_V$ of the vector fields $\mathcal{A}^0_\mu$, $A^\alpha_\mu$.
The associated field strengths $F^I_{\mu \nu}$, when combined with 
their duals $G_{I | \mu \nu}$ form a vector $(F^I_{\mu \nu}, G_{I | \mu \nu})$ 
which transforms linearly under $Sp(2n_V+2,\mathbb{R})$. The dual field strengths 
are dependent quantities,
which are defined as $G^{\pm}_{I|\mu \nu} = \partial \mathcal{L}/\partial
F^{\pm | I }_{\mu \nu}$, where $\mathcal{L}$ is the Lagrangian, and 
where $F^{\pm | I }_{\mu \nu}$ and $G^{\pm}_{I|\mu \nu}$
are the (anti-)selfdual parts of 
$F^{I }_{\mu \nu}$ and $G_{I|\mu \nu}$. We remark that the linear action 
of the symplectic group is obvious if one uses the gauge equivalence 
between $\mathcal{N}=2$ Poincar\'e supergravity with $n_V$ vector 
and $n_H$ hypermultiplets to $\mathcal{N}=2$ conformal supergravity
with $n_V+1$ vector and $n_H+1$ hypermultiplets. In the superconformal
setting $A^I_{\mu}$ are the vector fields of the  $n_V+1$ superconformal 
vector multiplets. The corresponding scalars $X^I$ allow a symplectically
covariant description of the scalar sector. In terms of the $X^I$ 
the prepotential is a holomorphic function $F(X)$ which is homogeneous
of degree 2, $F(\lambda X) = \lambda^2 F(X)$. Combining the scalars
$X^I$ with $F_I = \partial F/\partial X^I$ one obtains another symplectic
 vector $(X^I, F_I)$. The scalars $z^\alpha$ can be recovered as ratios
 $z^\alpha = X^\alpha/X^0$. The couplings between scalar and 
 vector fields are encoded in a complex matrix $\mathcal{N}_{IJ} = 
 \cR_{IJ} + i \I_{IJ}$, which can be expressed in terms of the
 prepotential. The kinetic terms for the scalar and vector 
 fields are positive definite if $g_{\alpha \bar{\beta}}$ is positive definite
 and if $\I_{IJ}$ is negative definite (in our convention). 

The Lagrangian for the bosonic degrees of the supergravity multiplet
and of $n_V$ vector multiplets takes the form
\begin{equation}
\label{4d_VM_Lagrangian}
\begin{aligned}
	L_{G + VM} =   \half &\star R_4 - {g}_{\alpha\bar{\beta}} (z, \bar{z}) dz^\alpha \wedge \star 
	d \bar{z}^{\bar{\beta}} - \frac{\lambda}{4} \I_{IJ} F^I \wedge \star F^J
	+ \frac{1}{4} \cR_{IJ} F^I \wedge F^J\;,
\end{aligned}
\end{equation}	
where $\lambda=-1$. 

We now turn to the modifications which occur if we change the supersymmetry 
algebra. In signature (1,3) we have the twisted algebra with R-symmetry
group $U(1,1)$. For this algebra the Lagrangian takes exactly the same form,
but with $\lambda=1$, that is, the signs of the kinetic terms for all vector
fields are flipped \cite{Cortes:2019mfa}.\footnote{This sign flip had already 
been observed in \cite{Sabra:2017xvx} by comparing the reductions 
of vector multiplets coupled to supergravity from signatures (1,4) and (2,3)
to signature (1,3). See  \cite{Cortes:2019mfa} for a detailed explanation how
this sign flip is related to the underlying R-symmetry groups.}
 While the scalar manifold remains the same (for a given
prepotential), we will use the notation SK$_\pm $= SK$_{\mp \lambda}$ to keep
track of the relative sign between scalar and vector fields. Note that 
SK$_+$ corresponds to the case with standard kinetic terms, $\lambda=-1$. 

Something more drastic happens in signatures (0,4) and (2,2), where 
special K\"ahler geometry is replaced by special para-K\"ahler geometry.
We will provide a concise summary and refer to the review \cite{LopesCardoso:2019mlj}
as well as the original papers  \cite{Cortes:2003zd,Cortes:2009cs,Cortes:2019mfa} for details.
Para-complex geometries are modelled on the para-complex numbers
(also called split complex numbers) in the same way as complex geometries
are modelled on the complex numbers. The para-complex numbers are
obtained by replacing the complex unit $i$, which satisfies $i^2=-1$ and 
$\bar{i}=-i$ by the para-complex unit $e$, which satisfies $e^2=1$ and
$\bar{e}=-e$. This allows one to define `para-analogues' of almost complex,
complex, Hermitian, K\"ahler and of affine and projective special K\"ahler geometry. 
For example, an almost para-complex structure $J$ on an even-dimensional
real manifold $\mathcal{M}$ is an endomorphism field $J\in \mbox{End}(T\mathcal{M})$
which satisfies $J^2= \mbox{Id}_{T\mathcal{M}}$ and has an equal number of
eigenvalues $\pm 1$. If $J$ is integrable, $\mathcal{M}$ admits local 
para-complex coordinates $z^i = x^i + e y^i$, and is a para-complex manifold.
Special para-K\"ahler (SPK) geometry is the para-analogue of special 
K\"ahler geometry. All usual formulae take the same form (assuming some care
in placing factors $e$), with the prepotential now a para-holomorphic function 
of para-complex scalar fields $z^\alpha$.

The change from complex to para-complex target geometry is 
reflected by the change in the abelian factor of the R-symmetry group. 
For special K\"ahler targets, the infinitesimal action of 
$U(1) \subset U(2)$ is given by multiplication by the complex structure $I$. 
Similarly the infinitesimal action of $SO(1,1)\subset U(1,1)$ is given 
by multiplication by the para-complex structure $J$ \cite{Cortes:2003zd}. 
Thus Table \ref{Tab:4d_susy}
tells us immediately that the vector multiplet geometry is SK for signature
(1,3) but SPK for signature (0,4) and (2,2). This can also be verified by 
explicit construction of the vector multiplet representations, which in addition
fixes the relative sign between the scalar and 
vector field terms \cite{Cortes:2003zd,Cortes:2019mfa}. As we will review
later, SPK geometry arises when reducing Euclidean IIA supergravity
on a Calabi-Yau threefold \cite{Sabra:2015tsa}. 
Note that if the scalar manifold is 
SPK, this relative sign does not really matter, that is 
we can take $\lambda=-1$ or $\lambda=1$,
because this sign can be flipped by a local field 
redefinition \cite{Cortes:2019mfa}. This reflects that in signatures
$(0,4)$ and $(2,2)$ the supersymmetry algebra is unique, whereas 
in signature $(1,3)$ there are two inequivalent supersymmetry 
algebras, whose vector multiplet representations are distinguished
by the relative sign between scalar and vector field terms. 
In Minkowski signature sign flips of the gauge kinetic term map 
solutions of one theory to solutions of the other. For planar 
Reissner-Nordstrom-like solutions, this defines a map which
exchanges the regions inside and outside horizons, and maps
cosmological to black hole solutions \cite{Gutowski:2020fzb}. 
In contrast, in Euclidean and neutral signature solutions with flipped
vector kinetic terms are related to one another by a field redefinition 
\cite{Sabra:2021ugi}.

\subsection{Hypermultiplets \label{HM}}

Hypermultiplets exist in all dimensions $D\leq 6$. 
Their field content is four real scalars and a doublet of spinors. 
The scalar geometry does not change under dimensional reduction. 
In Lorentz signature the scalar geometry is 
hyper-K\"ahler (HK) in the rigid case and quaternion-K\"ahler (QK) in the 
local case. A detailed review in conventions close to ours can be found
in \cite{Freedman:2012zz}.
In both cases the scalar manifold $\mathcal{N}$ 
carries the action of a quaternionic structure, which is 
spanned (at least locally) by three complex structures $I_i$, $i=1,2,3$, which 
satisfy the quaternionic algebra, that is they 
mutually anticommute and satisfy $I_i I_j = I_k$ for $i,j,k$ cyclic. 
Hypermultiplet scalars are charged under a non-abelian 
subgroup $SU(2)$ of the R-symmetry group, 
and the infinitesimal action of $SU(2)$ is
given by multiplication with 
the complex structures $I_i$. The corresponding finite action is given
by the unit quaternions, $a 1 + b I_1 + c I_2 + d I_3$, where
$a^2+b^2+c^2+d^2=1$, which form a group isomorphic 
to $SU(2)$. Three-dimensonal hypermultiplets can be obtained 
from four-dimensional vector multiplets by dimensional 
reduction. This induces a map between (generic) SK manifolds
and (non-generic) QK manifolds. This map is known as the 
c-map \cite{Cecotti:1988qn,Ferrara:1989ik}. The resulting QK manifolds contain the SK manifold
they are constructed from as a totally geodesic submanifold, and
the QK manifold is a group bundle over
an SK base.

Table \ref{Table:4d_3d} shows that in various four- and
three-dimensional signatures a factor $SU(2)$ of the R-symmetry
group is replaced by $SU(1,1)$ relative to the standard Lorentz 
signature algebra. This indicates that 
the quaternionic structure of the HM scalar manifold is 
replaced by a para-quaternionic structure.
The para-quaternions
(also called split quaternions) are obtained by replacing 
two of the three complex units by para-complex units. The
para-quaternionic algebra is isomorphic to the algebra 
$\mathbb{R}(2)$ of real 2 $\times$ 2 matrices, and the 
group of unit para-quaternions is isomorphic to $SU(1,1)$. 
The para-analogues of hyper-K\"ahler (HK) and quaternion-K\"ahler (QK)
geometry are called para-hyper-K\"ahler (PHK) and para-quaternion-K\"ahler
(PQK) geometry. We refer to 
\cite{Cortes:2005uq,Cortes:2015wca} and the review \cite{LopesCardoso:2019mlj}
for details. As we will discuss below, there are versions of the c-map which 
map SK and SPK manifolds to PQK manifolds.

One case where we expect that the hypermultiplet geometry is PQK 
is signature (0,3). This has been verified explicitly by dimensional
reduction from signature (1,3) to signature (0,3), which defines the temporal c-map,
and form signature (0,4) to signature (0,3), which defines the Euclidean c-map \cite{Cortes:2015wca}. 
More generally the results of \cite{Cortes:2015wca} imply the following: suppose
that $\mathcal{M}_{2n_V}$ is a (projective) SK or SPK manifold 
with coordinates $z^\alpha=x^\alpha + i_{\epsilon_1} y^\alpha$, 
metric $g_{\alpha \bar{\beta}}$, $\alpha, \beta = 
1, \ldots, n_V$ and vector coupling matrix $\mathcal{N}_{IJ} = \cR_{IJ} + i_{\epsilon_1}
\I_{IJ}$, where $\epsilon_1 = -1$, $i_{-1} = i$ for SK and $\epsilon_1 = 1$, 
$i_1=e$ for SPK. In the SK case we assume that $g_{\alpha \bar{\beta}}$ is
positive definite and that $\I_{IJ}$ is negative definite.\footnote{If this condition
is relaxed one obtains QK manifolds of indefinite signature. See \cite{Lledo:2006nr} 
for special geometry with indefinite signature SK and QK target spaces.}
Consider the two-parameter family of bosonic Lagrangians
for $n_H = n_V +1$ hypermultiplets,
\begin{equation}
\begin{aligned}
	L_{HM}^{(\epsilon_1, \epsilon_2)} 
	=  &- {g}_{\alpha \bar{\beta}} d z^\alpha \wedge \star d \bar{z}^{\bar{\beta}} - \frac{1}{4} d \varphi \wedge \star d \varphi \\
	&+\epsilon_1  e^{-2 \varphi} 
		\left[d \tilde{\phi} + \half \left(\zeta^I d \tilde{\zeta}_I - \tilde{\zeta}_I d\zeta^I \right) \right] \wedge \star \left[d \tilde{\phi} + \half \left(\zeta^I d \tilde{\zeta}_I - \tilde{\zeta}_I d\zeta^I \right) \right] \\
		&-  \frac{\epsilon_2 }{2} e^{- \varphi} \left[ {\I}_{IJ} d \zeta^I \wedge \star d\zeta^J -
		\epsilon_1  {\I}^{IJ} \left(d \tilde{\zeta}_I + {\cR}_{IK} d \zeta^K \right)
		\wedge \star \left(d \tilde{\zeta}_I + {\cR}_{IK} d \zeta^K \right) \right] \;,
\end{aligned}
\label{HM_master}
\end{equation}
where $\epsilon_2=\pm 1$ is a second parameter. It was shown in \cite{Cortes:2015wca},
that the resulting HM manifold $\mathcal{N}_{4n_H} = \mathcal{N}_{4n_V+4}$
is QK for $(\epsilon_1=-1,\epsilon_2=-1)$ 
and PQK for the other three cases. Moreover for $(\epsilon_1=-1,\epsilon_2=1)$
the PQK manifold is a group bundle over an SK base (the space parametrized by the complex
scalars $z^\alpha$), 
while for $(\epsilon_1=1,\epsilon_2=\pm 1)$ it is a group bundle over a SPK 
base (the space parametrized by the para-complex
scalars $z^\alpha$). Finally, the manifolds with $(\epsilon_{1}=1, \epsilon_2=\pm 1)$ 
are isometric (keeping the base manifold fixed). 
Thus there are three inequivalent cases: QK, PQK with an SK base
and PQK with an SPK base, see Table \ref{Tab:c-maps} for a summary.  If we need to empasize the base we will write
PQK$_{SK}$ or PQK$_{SPK}$. We remark that the c-maps $\mathcal{M}_{2n_V}
\rightarrow \mathcal{N}_{4n_V +4}$ are maps between S(P)K manifolds and 
Q(P)K manifolds, which are well defined on their own, that is without reference
to supermultiplets, Lagrangians and dimensional reduction. In particular the resulting QK/PQK manifolds 
are admissible (though non-generic)
HM target manifolds in all signatures for dimensions up to six, provided 
that they are compatible with the R-symmetry group.
We will see that QK/PQK manifolds of all of these types appear in 
type-II compactifications on Calabi-Yau threefolds.

\begin{table}
\begin{center}
\begin{tabular}{| l | l | l | l |} \hline
Domain & Image & Parameters  & c-map\\ \hline \hline
SK & QK & $\epsilon_1=-1, \epsilon_2=-1$ & spatial \\ \hline
SK & PQK$_{SK}$ &  $\epsilon_1=-1, \epsilon_2=1$ & temporal  \\ \hline
SPK & PQK$_{SPK}$ &  $\epsilon_1=1, \epsilon_2=\pm 1$ & Euclidean \\ \hline
\end{tabular}
\end{center}
\caption{As far as the scalar geometries are concerned, there are three distinct c-maps.
The parameters refer to the hypermultiplet `master Lagrangian' (\ref{HM_master}).
\label{Tab:c-maps}}
\end{table}

\subsection{Reduction to three dimensions}

Let us consider the dimensional reduction of $\mathcal{N}=2$ 
supergravity with $n_V$ vector multiplets and $n_H$ 
hypermultiplets to three dimensions. The field content and
scalar geometry of the HM sector does not change, 
while vector multiplets can be dualized into hypermultiplets
after reduction. Moreover, the bosonic degrees of freedom of the
supergravity multiplet, that is the metric and the 
graviphoton, give rise to an additional hypermultiplet, so
that we end up with three-dimensional $\mathcal{N}=4$
supergravity with $(n_V+1)+n_H$ hypermultiplets. Since 
the four-dimensional HMs play a passive role, we only need
to consider the bosonic Lagrangian (\ref{4d_VM_Lagrangian})
for gravity and $n_V$ vector multiplets. There are four
different starting points: signature (1,3) with either the standard
or twisted $\mathcal{N}=2$ algebra, signature (0,4) and signature (2,2).
The scalar geometry and relative signs are encoded in two
parameters: $\epsilon_1=\mp 1$ distinguishes between 
SK (signature (1,3)) and SPK (signatures (0,4), (2,2)), while
$\lambda=\pm 1$ encodes the relative sign between scalar
and vector terms. As mentioned earlier, the choice of this sign is only relevant
in signature (1,3), since in the other signatures it can be changed by 
a field redefinition. 
We introduce another parameter $\epsilon=\mp 1$, 
which distinguishes between spacelike reduction and timelike reduction. 
After the reduction, the Einstein-Hilbert term is non-dynamical, and the
local degrees of freedom of the four-dimensional metric reside in 
the KK-scalar $\varphi$ and the scalar $\tilde{\phi}$ which is dual 
to the KK-vector. The four-dimensional vector fields $A^I_\mu$ decompose into
scalars $\zeta^I$ and three-dimensional vector fields which we dualize
into scalars $\tilde{\zeta}_I$. Together with the (para-)\linebreak complex scalars
$z^\alpha$, this is the field content of $n_H+1$ hypermultiplets. 
The computation is the same as in \cite{Ferrara:1989ik} and 
\cite{Cortes:2015wca}, except that we now include the case $\lambda=+1$.
The  Lagrangian takes the form

\begin{equation}
\begin{aligned}
{\bf e}^{-1}  L_3 &= \half R_3 - \frac{1}{4}\partial_\mu \phi \partial^\mu \phi - g_{\alpha\bar{\beta}} \partial_\mu z^\alpha \partial^\mu \bar{z}^{\bar{\beta}} \\
 &+ \epsilon_1 e^{-2\phi} \left[ \partial^\rho \tilde{\phi} + \half \left(\zeta^I \partial^\rho \tilde{\zeta}_I 
 + \tilde{\zeta}_I \partial^\rho \zeta^I \right) \right] \left[ \partial_\rho \tilde{\phi} + \half \left(\zeta^I
 \partial_\rho \tilde{\zeta}_I - \tilde{\zeta}_\I \partial_\rho \zeta^I \right) \right] \\
 &+ \frac{\lambda \epsilon}{2} e^{-\phi} \left[\I_{IJ} 
 \partial_\mu\zeta^I \partial^\mu \zeta^J -\epsilon_1 \I^{IJ} 
	\left(\partial^\rho \tilde{\zeta}_I - \cR_{I K} \partial^\rho \zeta^K \right) 
	\left(\partial_\rho \tilde{\zeta}_J - \cR_{JL} \partial_\rho \zeta^L \right) \right] \;.
\end{aligned}
\end{equation}
By comparison to (\ref{HM_master}) we read off that $\lambda \epsilon = - \epsilon_2$.
All of these spaces are either QK, PQK$_{SK}$ or PQK$_{PSK}$. Starting with 
four theories in four dimensions, we have six different cases.
\begin{enumerate}
\item
Start with VMs in signature (1,3), with the standard $\mathcal{N}=2$ algebra, and
reduce over space, (1,3) $\rightarrow$ (1,2).
 Then $\epsilon_1=-1$ and $\epsilon=-1$, $\lambda =-1$ 
which implies $\epsilon_2=-1$. This is the standard (`spatial') c-map 
of \cite{Ferrara:1989ik}
which maps
$SK_+ \rightarrow QK$. 
\item
Start with VMs in signature (1,3), with the standard $\mathcal{N}=2$ algebra, and
reduce over time, (1,3) $\rightarrow$ (0,3). Then $\epsilon_1=-1$ and $\epsilon=1$, $\lambda =-1$ 
which implies $\epsilon_2=1$. This is the temporal c-map \cite{Cortes:2015wca}, which maps 
$SK_+ \rightarrow PQK_{SK}$. 
\item
Start with VMs in signature (0,4) and reduce over space, (0,4) $\rightarrow$ (0,3). 
Then $\epsilon_1=1$ and $\epsilon=-1, \lambda = \pm 1$ which implies $\epsilon_2=\pm 1$. 
This is the Euclidean c-map \cite{Cortes:2015wca}, 
which maps $SPK \rightarrow PQK_{SPK}$. 
\item
Start with VMs in signature (2,2) and reduce over time, (2,2) $\rightarrow$ (1,2). 
Then $\epsilon_1=1$ and $\epsilon=1, \lambda = \pm 1$ which implies $\epsilon_2=\pm 1$.
This works like the Euclidean c-map,  $SPK \rightarrow PQK_{SPK}$, but if we want
to emphasize the context of dimensional reduction, that is, that we reduce over time 
rather than space,  we will call it the neutral c-map.
\item 
Start with VMs in signature (1,3), with the twisted $\mathcal{N}=2$ algebra, and
reduce over space, (1,3) $\rightarrow$ (1,2).
 Then $\epsilon_1=-1$ and $\epsilon=-1$, $\lambda =1$ 
which implies $\epsilon_2=-1$. This maps SK to PQK with a SK base:
$SK_- \rightarrow PQK_{SK}$. Thus the sign flip between scalar and 
vector fields exchanges the roles of the spatial and temporal c-map. 
In the case at hand we obtain a PQK manifold from an SK manifold
through spatial reduction. While this works like the temporal c-map
as far as the scalar geometries are concerned, we will call this
the twisted spatial c-map if we want to emphasize the context of dimensional
reduction, that is, that we reduce over space, but start with flipped four-dimensional
gauge kinetic terms.
\item
Start with VMs in signature (1,3), with the twisted $\mathcal{N}=2$ algebra, and
reduce over time, (1,3) $\rightarrow$ (0,3).
 Then $\epsilon_1=-1$ and $\epsilon=1$, $\lambda =1$ 
which implies $\epsilon_2=1$. This maps SK to QK despite that 
we are reducing over time:
$SK_- \rightarrow QK$. While this works like the spatial c-map as
far as the manifolds are concerned, we will call this the twisted
temporal c-map if we need to emphasize the context of dimensional
reduction.
\end{enumerate}
See Table \ref{Tab:4d/3d} for a summary.

\begin{table}
\begin{center}
\begin{tabular}{| l | l | l | l | l |} \hline 
4d signature & Source & 3d signature & Image & c-map \\ \hline \hline
(0,4) & SPK & (0,3) & PQK$_{SPK}$ & Euclidean \\ \hline
(1,3) & SK$_+$ & (0,3) & PQK$_{SK}$ & temporal \\ \hline
(1,3) &  SK$_-$ & (0,3) & QK & twisted temporal $\cong$ spatial \\ \hline
(1,3) & SK$_+$ & (1,2) & QK & spatial  \\ \hline
(1,3) &  SK$_-$ & (1,2) & PQK$_{SK}$ & twisted spatial $\cong$ temporal \\ \hline
(2,2) & SPK & (1,2) &  PQK$_{SPK}$ & neutral $\cong$ Euclidean \\ \hline
\end{tabular}
\end{center}
\caption{When reducing four-dimensional vector multiplets to three dimensions, there
are six distinct cases, although there are only three distinct types of hypermultiplet 
manifolds that arise from the construction. \label{Tab:4d/3d}}
\end{table}

If we start with a theory of $n_V$ vector and $n_H$ hypermultiplets in four
dimensions, with scalar manifold $\mathcal{M}_{2n_V} \times \tilde{\mathcal{N}}_{4n_H}$,
reduction to three dimensions leads us to a theory with 
$(n_V+1)+n_H$ hypermultiplets, where the two hypermultiplet manifolds
form a direct product:
\[
\mathcal{M}_{2n_V} \times \tilde{\mathcal{N}}_{4n_H} \rightarrow 
\mathcal{N}_{4n_V + 4}\times  \tilde{\mathcal{N}}_{4n_H} \;.
\]
If both factors are `in the image of the c-map', 
the three-dimensional theory can be lifted to a different four-dimensional 
theory with $n'_V = n_H-1$ vector multiplets and $n'_H = n_V+1$ 
hypermultiplets. 
\begin{equation}
\mathcal{M}_{2n_V} \times \tilde{\mathcal{N}}_{4n_H} \rightarrow 
\tilde{\mathcal{N}}_{4n_V + 4} \times \tilde{\mathcal{N}}_{4n_H}
\leftarrow
\tilde{\mathcal{N}}_{4n_V + 4} \times \tilde{\mathcal{M}}_{2 n_H - 2} =
\tilde{\mathcal{N}}_{4n'_H} \times \tilde{\mathcal{M}}_{2n'_V} \;.
\end{equation}

In the context of string theory, the relations between the four-dimensional
theories are T-dualities, which we call spacelike, timelike and mixed
depending on how they combine spacelike/timelike reduction with 
spacelike/timelike oxidation.
Which T-dualities exist depends
on the details of the HM sectors of the four-dimensional theories. 
Therefore we will now consider the Calabi-Yau compactifications of 
type-II theories in signature (0,10), (1,9) and (2,8), which give rise
to four-dimensional $\mathcal{N}=2$ theories in signatures
(0,4), (1,3) and (2,2).

\section{Ten-dimensonal type-II string theories \label{sec:type-II}}

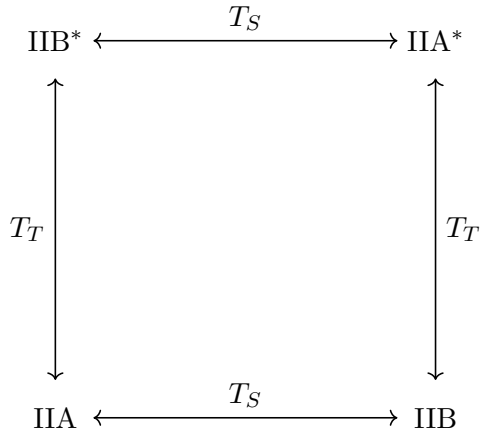
\begin{figure}[!h]
	\centering
	\begin{tikzpicture}	
		\node at (0,0) {IIA};
		\node at (5,5) {IIA$^*$};
		\node at (0,5) {IIB$^*$};
		\node at (5,0) {IIB};
		\draw[semithick,<->] (0.5,0) -- (4.5,0) node [midway, above] {$T_S$};
		\draw[semithick,<->] (0.5,5) -- (4.5,5) node [midway, above] {$T_S$};
		\draw[semithick,<->] (0,0.5) -- (0,4.5) node [midway, left] {$T_T$};
		\draw[semithick,<->] (5,0.5) -- (5,4.5) node [midway, right] {$T_T$};
	\end{tikzpicture}
	\caption{Diagram showing the relationship between II and II$^*$ theories, where $T_S (T_T)$ denotes a spacelike (timelike) T-duality.}
	\label{IImap}
\end{figure}

As is well known, type-IIA and type-IIB string theory are related by 
T-duality. When admitting timelike T-duality, one obtains two further
theories, dubbed type-IIA$^*$/IIB$^*$, as summarized in Figure \ref{IImap} \cite{Hull:1998vg}.

The difference between type-IIA and type-IIA$^*$ and between type-IIB 
and type-IIB$^*$ lies in certain phase factors,
which at the level of the effective supergravity Lagrangian manifest themselves in 
sign flips of the kinetic terms of the R-R fields, as well as factors of powers of $i$ 
in the fermionic terms. For type-IIB/IIB$^*$ the scalar manifolds are different, 
namely $SL(2,\mathbb{R})/SO(2)$ for type-IIB and $SL(2,\mathbb{R})/SO(1,1)$ for type-IIB$^*$. 
It was observed that supersymmetry is realized in type-II$^*$ theories in a modified,
twisted form, which can be interpreted as a generalized $O(p,q)$ Majorana
condition \cite{Hull:1998ym}.  In \cite{Gall:2021tiu} the R-symmetry groups for
supersymmetry algebras in arbitrary dimension and signature were classified,
which allows to put this observation into a wider context. It was found that 
in certain signatures there exist several non-isomorphic supersymmetry algebras
with the same number of supercharges (and, where applicable the same chirality
properties), whose R-symmetry groups are different real forms of the same 
complex Lie group. For example, in signature (1,9), chiral supersymmetry 
algebras with $\mathcal{N}$ left-moving (or right-moving) supercharges
are real forms of a complex supersymmetry algebra with R-symmetry 
group $O(\mathcal{N},\mathbb{C})$. Real supersymmetry algebras are
obtained by imposing $O(p,q)$ Majorana
conditions,  with $p+q=\mathcal{N}$, which leads to real supersymmetry algebras
with R-symmetry group $O(p,q)$. 
For ten-dimensional
chiral supersymmetry algebras with 32 real supercharges the two possible cases
are $O(2)$ and $O(1,1)$ which correspond to type-IIB/IIB$^*$. There also are
two inequivalent non-chiral algebras, which have the same discrete R-symmetry 
group but differ by a relative sign in the reality condition imposed on left- and 
right-moving supercharges, corresponding to type-IIA/IIA$^*$. 
Similarly, $\mathcal{N}$-extended supersymmetry algebras in four-dimensions
are real forms of a complex supersymmetry algebra with R-symmetry $GL(\mathcal{N},\mathbb{C})$ 
and the reality conditions defining real supersymmetry algebras in signature
(1,3) lead to R-symmetry groups of the form $U(p,q)$, $p+q = \mathcal{N}$. 
For $\mathcal{N}=2$ the two possibilities are $U(2)$ and $U(1,1)$. 
For completeness we note that for $\mathcal{N}=2$ the reality conditions 
defining real supersymmetry algebras in signatures (0,4) and (2,2) 
lead to unique algebras with R-symmetry $U^*(2)$ and $GL(2, \mathbb{R})$, 
respectively, see \cite{Gall:2021tiu} for details.

All type-II theories have the same NS-NS sector which consists of the graviton 
$G_{MN}$, Kalb-Ramond field $B_{MN}$ and dilaton $\Phi$. The R-R sector
of type-IIA/IIA$^*$ contains a one form $C_1$ and a three-form $C_3$ 
while the R-R sector of type-IIB/IIB$^*$ contains a zero-form $C_0$, 
a two-form $C_2$ and a four-form $C_4$ whose field strength is
self-dual or anti-self-dual, $*G_5 = \pm G_5$.\footnote{We will specify our choice of
sign below.} 
 The difference between the bosonic actions
of type-II and type-II$^*$ is a sign flip of the kinetic terms for all fields in 
R-R sector, see Tables \ref{Tab:IIA} and \ref{Tab:IIB}.\footnote{The information for type-IIA
is taken from Table 1 in \cite{Hull:1998ym}. Note that their notation for signature
is $(s,t)$, where $s$ corresponds to positive eigenvalues of the metric, and $t$ 
to negative eigenvalues of the metric, while we use $(t,s)$. }

\begin{table}
\begin{center}
\begin{tabular}{| l | l | l | l | l | l |} \hline
Type & $G_{MN}$ & $B_{MN}$ & $\Phi$ &  $C_1$ & $C_3$ \\ \hline \hline
IIA$_{(1,9)}$ & + & + & + & + & + \\ \hline
IIA$^*_{(1,9)}$ & + & + & + & $-$ & $-$ \\ \hline
IIA$_{(0,10)}$ & + & $-$ & + &  $-$ & + \\ \hline
IIA$_{(2,8)}$ & + & $-$ & + &  + & $-$ \\ \hline
\end{tabular}
\end{center}
\caption{Relative signs for kinetic terms in ten-dimensional type-IIA theories. 
A + sign corresponds to a standard kinetic term in Lorentz signature, thus
discarding the overall $-$ sign with which these terms appear in the action 
when using the mostly plus convention 
for the metric.
\label{Tab:IIA}}
\end{table}

\begin{table}
\vspace{1cm}
\begin{center}
\begin{tabular}{| l | l | l | l | l | l | l | } \hline
Type & $G_{MN}$ & $B_{MN}$ & $\Phi$ &  $C_0$ & $C_2$ & $C_4$ \\ \hline \hline
IIB$_{(1,9)}$ & + & + & + & + & + & + \\ \hline
IIB$^*_{(1,9)}$ & + & + & + & $-$ & $-$ & $-$ \\ \hline
IIB'$_{(1,9)}$ & + & $-$ & + & $-$ & + & $-$ \\ \hline
\end{tabular}
\end{center}
\caption{Relative signs for kinetic terms in ten-dimensional type-IIB theories. A + indicates
the standard sign for a theory in Lorentz signature using the mostly plus convention.\label{Tab:IIB}}
\end{table}

The bosonic actions for type-IIA/IIA$^*$ take the form \cite{Hull:1998ym}
\begin{equation}
\label{Action1,9IIA}
S_{(1,9)}^{IIA/IIA^*} = \int d^{10} x \sqrt{|G|} e^{-2\Phi} \left(R + 4 (\partial \Phi)^2 - H^2 + \lambda G_2^2 
+ \lambda G_4^2 \right) + \cdots
\end{equation}
where we omitted the Chern-Simons terms, and where $H,G_2,G_4$ are the field strength
of $B,C_1,C_3$, respectively. Type-IIA corresponds to $\lambda=-1$ while type-IIA$^*$ 
corresponds to $\lambda=1$. Taking into account that we use the mostly plus
convention for the metric, this means that all bosonic fields have positive kinetic energy
for $\lambda=-1$. In Table \ref{Tab:IIA} this corresponds to a row where all entries 
are $+$, that is we discard the overall minus sign that these terms have in the action.
Generally, in this and other tables, we record sign flips relative to standard kinetic terms
in Lorentz signature, which correspond to a row with only $+$ signs. Note that the 
kinetic term of the dilaton in the action (\ref{Action1,9IIA})
has a $+$ sign, since we are in the string frame. When 
going to the Einstein frame, this sign flips, showing that the dilaton has positive kinetic
energy. 

For type-IIB/IIB$^*$ there is no simple covariant action, since the five-form field strength
is self-dual. However one can use a pseudo-action, whose variation gives the field
equation except the self-duality condition $G_5=\pm *G_5$,  which is then imposed 
by hand \cite{Hull:1998ym}:\footnote{The sign in the self-duality relation 
is correlated with the sign of terms
which we have not displayed (Chern-Simons and fermionic terms), see
for example \cite{Ortin} for a general discussion.  Full bosonic type-II
Lagrangians, which however use a different notation and normalization for the bosonic fields, 
can be found in \cite{Dijkgraaf:2016lym}. In their conventions the sign is correlated
with whether the worldvolume theories of fundamental strings and D-strings are
Lorentzian or Euclidean, resulting in a $(+)$-sign for type-IIB and a $(-)$-sign for 
type-IIB$^*$/IIB'.} 
\begin{equation}
S_{(1,9)}^{IIB/IIB^*} = \int d^{10} x \sqrt{|G|}  e^{-2\Phi} \left(R + 4 (\partial \Phi)^2 - H^2 +
\lambda G_1^2 + \lambda G_3^2 + \lambda G_5^2 \right) + \cdots
\end{equation}
where $G_{p+1} = dC_p + \cdots$ are the field strength and where again we only display
the Maxwell-like terms. Type-IIB corresponds to $\lambda=-1$, where all kinetic terms
have their standard sign, while type-IIB$^*$ exhibits a sign flip for all R-R fields. 

Type-II string theories exist for all ten-dimensional signatures. We will use
the notation type-II$_{(t,s)}$ for a theory where the metric has $t$ negative
and $s$ positive eigenvalues. Since we prefer the mostly plus convention 
for Lorentz signature, we will usually refer to the the $t$ directions as timelike
and the $s$ directions as spacelike. Theories where $t$ and $s$ are exchanged
have been shown to be equivalent \cite{Hull:1998ym}.
The unique theories
in signatures (0,10) and (2,8) are IIA theories, denoted IIA$_{(0,10)}$ and IIA$_{(2,8)}$.
These theories are non-chiral and have 
the same R-R sector as type-IIA$_{(1,9)}$. Their actions have the same structure
as the type-IIA$_{(1,9)}$ action, but with some sign flips for the Maxwell-like terms
 \cite{Hull:1998ym},  which are listed in Table \ref{Tab:IIA}. In both signatures the 
 $B$-field has a flipped kinetic term, while in the R-R sector either $C_1$ or
 $C_3$ has a sign flip:
 \begin{eqnarray}
 S_{(0,10)}^{IIA} &=&  \int d^{10} x \sqrt{|G|} e^{-2\Phi} \left(R + 4 (\partial \Phi)^2 + H^2 + G_2^2 
- G_4^2 \right) + \cdots  \;, \\
 S_{(2,8)}^{IIA} &=&  \int d^{10} x \sqrt{|G|} e^{-2\Phi} \left(R + 4 (\partial \Phi)^2 +  H^2 - G_2^2 
+ G_4^2 \right) + \cdots  \;.
\end{eqnarray}

 Type-II string theories in different signatures are related by what we call mixed
 T-dualities, that is T-dualities which combine a spacelike/timelike reduction 
 with a timelike/spacelike oxidation (lifting). For this to work one needs to make
 use of the S-dual of the type-IIB$^*$ theory, which is called type-IIB'. 
 As shown in \cite{Hull:1998ym} Buscher T-duality along an isometric 
 direction $X^\sharp$ in the target space of the worldsheet sigma model
 preserves the sign of the term $G_{\sharp \sharp} \partial_\alpha X^\sharp \partial^\alpha X^\sharp$
 if the worldsheet theory has  Lorentzian signature, but reverses it if the 
 worldsheet theory has Euclidean signature. 
 In the type-IIB$^*$-theory, D-branes are replaced by E-branes, which have a Euclidean
 worldvolume. If one applies S-duality, fundamental IIB$^*$-strings and E-strings are 
 exchanged, so that in the resulting IIB'-theory fundamental strings have a Euclidean
 worldvolume. The type-IIA$_{(0,10)}$ and type-IIA$_{(2,8)}$ theories are then obtained
 from the type-IIB'$_{(1,9)}$ theory
 by T-dualities which involve a timelike/spacelike reduction combined by a 
 spacelike/timelike oxidation
  \cite{Hull:1998ym}. At the supergravity level, S-duality exchanges the $B$-field
 and the R-R two-form $C_2$, resulting in the sign flips recorded in Table \ref{Tab:IIB}, 
 \begin{equation}
 S_{(1,9)}^{IIB'} = \int d^{10} x \sqrt{|G|}  e^{-2\Phi} \left(R + 4 (\partial \Phi)^2 + H^2 +
 G_1^2 -  G_3^2 +  G_5^2 \right) + \cdots
 \end{equation}

 Note that while $G_5$ is an S-duality singlet, $\Phi$ and $C_0$ parametrize
 the indefinite signature coset space $SL(2,\mathbb{R})/SO(1,1)$ on which S-duality acts
 non-linearly. 

\section{Type-II Calabi-Yau compactifications}

By compactification of type-II string theories one obtains $\mathcal{N}=2$ supergravity
coupled to vector and hypermultiplets. 
Since we always compactify six spatial dimensions to go from signatures 
(0,10), (1,9), (2,8) to signatures (0,4), (1,3), (2,2), 
the only essential difference in these reductions 
is between type-IIA and type-IIB, which are distinguished, as far as bosonic
degrees of freedom are concerned, by the field content of their R-R sectors. 
Otherwise the bosonic type-IIA/IIA$^*$ actions only differ from one
another by relative sign flips that one has to follow through, and the same
applies to the type-IIB/IIB$^*$/IIB' theories. 
Since the mechanics of Calabi-Yau compactifications is well known
from the standard cases of IIA$_{(1,9)}$ \cite{Bodner:1990zm}
and IIB$_{(1,9)}$ \cite{Bohm:1999uk}, we will not go through
the computational details but highlight how the ten-dimensional sign flips modify the resulting 
four-dimensional actions. More details are given in the Appendix. The reduction of the Euclidean 
IIA$_{(0,10)}$ theory was worked out in detail in  \cite{Sabra:2015tsa}.

\subsection{Type-IIA Calabi-Yau compactifications}

We start with type-IIA theories, where we have the cases
type-IIA$_{(1,9)}$, IIA$^*_{(1,9)}$, IIA$_{(0,10)}$ and IIA$_{(2,8)}$. We first consider aspects
which work the same in all cases.

{\bf The metric.}
In a general real six-fold compactification,  the massless four-dimensional
fields resulting from the reduction of the metric $G_{MN}$ are  the four-dimensional
metric $g_{\mu \nu}$, vector fields, and scalar fields. Massless vector fields are on one-to-one
correspondence with Killing vector fields, and since CY3-folds (Calabi-Yau threefolds)
do not have isometries, 
there are no massless vectors in our case. The massless scalar fields are in one-to-one with deformations
of the six-fold metric which preserve Ricci-flatness. For CY3-folds these deformations are, due to 
the existence of a holomorphic (3,0)-form, in one-to-one correspondence with the deformations
of the complex structure and of the (real) K\"ahler form. This gives rise to $h^{2,1}$ complex
scalars $z^\alpha$, $\alpha = 1, \ldots, h^{2,1}$ and $h^{1,1}$ real scalars $y^A$, $A=1, \ldots, h^{1,1}$,
where $h^{i,j}$ are the Hodge numbers of the CY3-fold.

{\bf $p$-form fields.} A ten-dimensional $p$-form decomposes into products of 
four-dimensional $p'$-forms and six-dimensional $p''$-forms, where $p'+p''=p$. 
Massless $p'$-forms are in one-to-one correspondence with harmonic $p''$-forms, 
which are counted by the Betti-numbers $b_{p''}$ of the compact space. For a CY3-fold
the Betti numbers are related to the Hodge numbers by
$b_{p''} = \sum_{i+j=p''} h^{i,j}$. Moreover, for a CY3-fold
\[
h^{0,0} = h^{3,0} = h^{0,3} = h^{3,3} = 1 \;,\;\;\;
h^{1,0}=h^{0,1} = h^{3,2}=h^{2,3} =0 \;,\;\;
h^{3,0}=h^{0,3} = 1 \;,
\]
so that the only numbers that vary between CY3s are $h^{1,1}=h^{2,2}\geq 1$ and 
$h^{1,2}=h^{2,1}\geq 0$.

{\bf The B-field.}
The $B$-field $B_{MN}$ gives rise to $h^{1,1}$ real scalars $x^A$, as well as
a four-dimensional $B$-field, which we dualize into a scalar
$\tilde{\phi}$.

{\bf The dilaton.}
The ten-dimensional dilaton $\Phi$ gives rise to a four-dimensional scalar $\varphi$, which differs
from $\Phi$ by a field-redefinition. Essentially, one absorbs a factor proportional to the volume of
the internal space, in order that the four-dimensional action acquires standard form.

{\bf The R-R sector.} The R-R one-form $C_M$ gives rise to a vector $\mathcal{A}^0_\mu$.
The R-R three-form $C_{MNP}$ gives rise to $h^{1,1}$ vectors $\mathcal{A}^A_\mu$ and $2h^{2,1}+2$ real
scalars $\zeta^I, \tilde{\zeta}_I$, $I=0, 1, \ldots, h^{2,1}$. 
The massless fields originating from the NS-NS sector are summarized 
in Table \ref{Tab:Massless_NS-NS}, those from the R-R sector 
in Table \ref{Tab:Massless_R-R-IIA}.

\begin{table}
\begin{center}
\begin{tabular}{| l | l| l| } \hline
  10d & 4d & \\ \hline \hline
  $G_{MN}$ & $g_{\mu \nu}$ & Metric \\
           &$ z^\alpha$ & Complex structure moduli \\
           & $y^A$ & (Real) K\"ahler moduli \\ \hline
  $B_{MN}$ & $b_{\mu \nu} \sim \tilde{\phi}$ & Universal axion \\
           & $x^A$ & $h^{1,1}$ real scalars \\ \hline
  $\Phi$ & $\varphi$ & Dilaton \\ \hline
 \end{tabular}
 \end{center}
 \caption{Massless fields in type-II Calabi-Yau compactifications, NS-NS sector.
 \label{Tab:Massless_NS-NS}}
\end{table}

\begin{table}
\begin{center}
 \begin{tabular}{| l | l| l| } \hline
   10d & 4d & \\ \hline \hline 
   $C_M$ & $\mathcal{A}^0_\mu$ & vector \\ \hline
   $C_{MNP}$ & $C_{\mu np} \sim \mathcal{A}^A_\mu$ & $h^{1,1}$ vectors \\ 
    & $ C_{mnp} \sim \zeta^I , \tilde{\zeta}_I$ & $2h^{2,1}+2$ scalars
   \\ \hline
   \end{tabular}
   \end{center}
   \caption{Massless fields in type-IIA Calabi-Yau compactifications, R-R sector.
   \label{Tab:Massless_R-R-IIA}}
\end{table}

Collecting all these fields, this is the bosonic field content of the $\mathcal{N}=2$ Poincar\'e supergravity
multiplet, $(g_{\mu \nu}, \mathcal{A}^0_\mu)$, of $n_V = h^{1,1}$ vector multiplets 
$(y^A, x^A, \mathcal{A}^A_\mu)$, and of $n_H=h^{2,1}+1$ hypermultiplets
$(z^\alpha, \varphi, \tilde{\phi}, \zeta^I, \tilde{\zeta}_I)$. The signs of the kinetic terms of the scalar 
fields can be inferred from those of the higher-dimensional ones, and are listed in 
Table \ref{Tab:IIACY3}.

\begin{table}
\begin{center}
\begin{tabular}{| l || l | l || l | l | l | l | l || l | l |} \hline
 & $y^A$ & $x^A$ & $z^\alpha$ & $\varphi$ & $\tilde{\phi}$ & $\zeta^I$ & $\tilde{\zeta}_I$ & 
 $\mathcal{A}^0_\mu$ & $\mathcal{A}^A_{\mu}$  \\  \hline \hline
IIA$_{(1,9)}$ & + & + & + & + & + & + & + & + & +  \\ \hline
IIA$^*_{(1,9)}$ & + & + & + & + & + & $-$ & $-$ & $-$ & $-$ \\ \hline
IIA$_{(0,10)}$ & + & $-$ & + & + & + & + & + & $-$ & +  \\ \hline
IIA$_{(2,8)}$ & + & $-$ & + & + & + & $-$ & $-$ & + & $-$ \\ \hline
\end{tabular}
\end{center}
\caption{Signs of the kinetic terms for scalar and vector 
fields resulting from type-IIA CY3 compactifications. A $+$ 
indicates a standard kinetic term. The fields $y^A,x^A$ are the vector multiplet scalars.\label{Tab:IIACY3}}
\end{table}

Signs are taken relative to the standard IIA$_{(1,9)}$ theory, where all kinetic terms
have the standard sign, denoted +. In IIA$^*_{(1,9)}$ half of the signs in the HM
sector are flipped, so that the HM scalar manifold has neutral signature. In the 
Euclidean IIA$_{(0,10)}$ theory only the signs of the scalars $x^A$ which descend from 
the $B$-field are flipped, which gives the VM manifold neutral signature. 
Finally in the IIA$_{(2,8)}$ case, we have 
signs flips for $x^A, \zeta^I, \tilde{\zeta}_I$, so that both VM and HM scalar manifold
have neutral signature.\footnote{Note that in both cases the sign flip of  the four-dimensional
$B$-field $b_{\mu \nu}$ is
compensated by a second sign flip when we dualize this two-form into the
scalar $\tilde{\phi}$. Dualization flips the sign of the kinetic term for a $p$-form 
field if and only if the metric has an even number of negative eigenvalues.}

The ten-dimensional sign flips also affect the four-dimensional vector kinetic terms. 
For type-IIA$^*_{(1,9)}$ the signs of all vector kinetic terms are flipped, whereas
for type-IIA$_{(0,10)}$ only the sign of $\mathcal{A}^0_\mu$ is flipped, while
for type-IIA$_{(2,8)}$ only the signs of $\mathcal{A}^A_\mu$ are flipped. 
Thus the vector kinetic terms have signatures $(+)^{h^{1,1}+1}$, 
$(-)^{h^{1,1}+1}$, $(+)^{h_{1,1}}(-)$, and $(+)(-)^{h^{1,1}}$, respectively.

The interactions of these fields are encoded in certain coupling matrices that one 
obtains when performing the dimensional reduction. For a four-dimensional 
$\mathcal{N}=2$ theory these coupling matrices can be interpreted as geometrical data
on the scalar manifolds $\mathcal{M}_{2 h^{1,1}}$ of the vector and 
$\mathcal{N}_{4 h^{2,1} +4}$ of hypermultiplets, which have real dimensions
$2h^{1,1}$ and $4 h^{2,1}+4$, respectively. These are the geometries that we
have reviewed in the previous section. 
At this point the sign flips become relevant
since they determine the signatures of the metrics of $\mathcal{M}_{2 h^{1,1}}$
and $\mathcal{N}_{4 h^{2,1} +4}$.

\subsubsection{The vector multiplet sector}

Let us first consider the vector multiplet scalars $y^A$ and $x^A$. 
In signature
$(1,9)$ their kinetic terms come with same sign, and the manifold $\mathcal{M}_{2 h^{1,1}}$
can be shown to be a complex manifold. The real scalars $y^A$ and $x^A$ can be combined into
complex scalars $z^A = y^A + i x^A$, which provide holomorphic coordinates 
for $\mathcal{M}_{2 h^{1,1}}$. The scalar fields $y^A$ parametrize the  moduli 
space of real K\"ahler forms $J$ on the CY3, while $x^A$ parametrize the deformations
of the internal components of the $B$-field, which corresponds to a harmonic (1,1)-form on 
the CY3. The combined moduli space parametrized by $z^A$ can be viewed as a
complexification of the real moduli space of K\"ahler forms, and is usually just 
called the K\"ahler moduli space. This space carries 
itself a K\"ahler metric $g_{A\bar{B}}(z,\bar{z})$, which appears in the four-dimensional
action as the generalized kinetic term (sigma model) of the scalars $z^A$, 
that is $\mathcal{L} \sim g_{A\bar{B}}(z,\bar{z}) \partial_\mu z^A \partial^\mu \bar{z}^{\bar{B}}$.
Moreover, this K\"ahler metric is not generic, but special, because its K\"ahler potential
$K(z,\bar{z})$ can be obtained from a holomorphic prepotential $\mathcal{F}(z)$. 
Thus $g_{A\bar{B}}(z,\bar{z})$ is a (projective) special K\"ahler metric or SK metric
for short.

Let us next look at the four-dimensional vector fields, still restricting ourselves
to signature (1,9). We have obtained $h^{1,1} +1$ vector fields, of which one, $\mathcal{A}^0_\mu$,
belongs to the supergravity multiplet and is called the graviphoton, while the others, 
$\mathcal{A}^A_{\mu}$, belong to the $h^{1,1}$ vector multiplets. We denote
the corresponding field strength by $\mathcal{F}^0_{\mu \nu}$ and $\mathcal{F}^A_{\mu \nu}$. 
As explained before, the vector fields can be rearranged into linear combinations $A^\Sigma_\mu$, 
$\Sigma=0, \ldots, n_V=h^{1,1}$ so that the field strength $F^\Sigma_{\mu \nu}$
together with their duals $G_{\Sigma|\mu \nu}$ form a symplectic vector. By carrying
out the reduction explicitly, one finds that the couplings between scalars and vectors
are encoded by the complex coupling matrix $\mathcal{N}_{\Sigma \Lambda} 
= \cR_{\Sigma \Lambda} + i \I_{\Sigma \Lambda}$, which depends on the scalars
$z^A$ through the prepotential $\mathcal{F}(z^A)$.

The resulting bosonic Lagrangian for the supergravity multiplet and $n_V$ vector multiplets
has the form (\ref{4d_VM_Lagrangian}), and the only difference 
between type-IIA and type-IIA$^*$ is the overall sign flip for the vector fields
$A^\Sigma_\mu$. Since we use a convention where $\I_{\Sigma \Lambda}$
is negative definite, type-IIA corresponds to $\lambda=-1$, while
type-IIA$^*$ corresponds to $\lambda=1$:
\begin{equation}
\label{4d_VM_Lagrangian_IIA/IIA-star}
\begin{aligned}
	L^{(1,3)IIA/IIA^*}_{G + VM} =   \half &\star R_4 - \bar{g}_{A\bar{B}} (z, \bar{z}) dz^A \wedge \star d \bar{z}^B - \frac{\lambda}{4} \I_{\Sigma \Lambda} F^\Sigma \wedge \star F^\Lambda 
	+ \frac{1}{4} \cR_{\Sigma \Lambda} F^\Sigma \wedge F^\Lambda\;,
\end{aligned}
\end{equation}	
where $A,B = 1, \ldots, n_V=h^{1,1}$ and $\Lambda, \Sigma = 0, \ldots, n_V = h^{1,1}$. 
For $\lambda=-1$ this is the standard result of \cite{Bodner:1990zm}. For 
the type-IIA$^*$ the sign flips in the ten-dimensional Lagrangian induce a sign flip
in the four-dimensional Maxwell term.

In signatures
$(0,10)$ and $(2,8)$ Table \ref{Tab:IIACY3} shows that
the metric of $\mathcal{M}_{2 h^{1,1}}$ has neutral signature. The four-dimensional
$\mathcal{N}=2$ supersymmetry algebra requires SPK geometry for the vector multiplets
in these cases.
The case (0,10) $\rightarrow$ (0,4) has been worked out in full detail in  \cite{Sabra:2015tsa}.
As far as the vector multiplet sector is concerned, the only difference between this
and the case (2,8) $\rightarrow$ (2,2) is an overall sign flip of the Maxwell term. 
Note that  in both cases the vector kinetic terms have Lorentz signature and therefore are indefinite. 
As mentioned before, the overall sign of the Maxwell term is conventional in the sense that it
can be flipped by a field redefinition. Therefore we can take either value of
$\lambda=\pm 1$ in the following Lagrangian:

\begin{equation}
\label{4d_VM_Lagrangian_para}
\begin{aligned}
	L^{(0,4),(2,2)}_{G + VM} =   \half &\star R_4 - \bar{g}_{A\bar{B}} (z, \bar{z}) dz^A \wedge \star d \bar{z}^B + \frac{\lambda}{4} \I_{\Sigma \Lambda} F^\Sigma \wedge \star F^\Lambda 
	+ \frac{1}{4} \cR_{\Sigma \Lambda} F^\Sigma \wedge F^\Lambda\;. 
\end{aligned}
\end{equation}	
Compared to (\ref{4d_VM_Lagrangian_IIA/IIA-star}) the scalar geometry is now SPK and the
scalar fields $z^A$ are para-complex fields.
The couplings $\bar{g}_{A\bar{B}}$,  $\I_{\Sigma \Lambda}$ and $\cR_{\Sigma \Lambda}$
are determined by the standard formulae of special geometry, but using a para-holomorphic
instead of a holomorphic prepotential, see  \cite{Cortes:2003zd,Cortes:2009cs,Sabra:2015tsa} 
for details.

\subsubsection{The hypermultiplet sector}

The scalars $z^\alpha$, $\alpha=1, \ldots, h^{2,1}$ parametrize the deformations of the complex
structure of the CY3 metric. They provide coordinates on a special K\"ahler submanifold of the
hypermultiplet manifold, with metric $g_{\alpha \bar{\beta}}(z,\bar{z})$ and 
prepotential $\mathcal{F}(z^\alpha)$. 

The additional scalars $\varphi$ (dilaton), $\tilde{\phi}$ (axion) and 
$\zeta^I, \tilde{\zeta}_I$, $I=0, \ldots h^{2,1}$ (R-R scalars) extend this SK manifold
either to a quaternion-K\"ahler  manifold (QK manifold)
or to a para-quaternion-K\"ahler  manifold (PQK manifold). Which case is realized depends
on the signs of the kinetic terms of the R-R scalars. 
The HM Lagrangian takes the form
\begin{equation}
\label{HM_IIA}
\begin{aligned}
	L_{HM}^{IIA} =  &- \tilde{G}_{\alpha \bar{\beta}} d z^\alpha \wedge \star d \bar{z}^{\bar{\beta}} - \frac{1}{4} d \varphi \wedge \star d \varphi \\
	&- e^{-2 \varphi} 
		\left[d \tilde{\phi} + \half \left(\zeta^I d \tilde{\zeta}_I - \tilde{\zeta}_I d\zeta^I \right) \right] \wedge \star \left[d \tilde{\phi} + \half \left(\zeta^I d \tilde{\zeta}_I - \tilde{\zeta}_I d\zeta^I \right) \right] \\
		&- \frac{\lambda}{2} e^{- \varphi} \left[ {\I}_{IJ} d \zeta^I \wedge \star d\zeta^J 
		+ {\I}^{IJ} \left(d \tilde{\zeta}_I + {\cR}_{IK} d \zeta^K \right)
		\wedge \star \left(d \tilde{\zeta}_I + {\cR}_{IK} d \zeta^K \right) \right] \;,
\end{aligned}
%\label{4d_HM}
\end{equation}
where $\lambda=-1$ for type-IIA$_{(1,9)}$ \cite{Bodner:1990zm} and 
type-IIA$_{(0,10)}$  \cite{Sabra:2015tsa} and $\lambda = 1$ for 
type-IIA$^*_{(1,9)}$ and type-IIA$_{(2,8)}$. 
The coupling matrices $\I_{IJ}$ and $\cR_{IJ}$ depend on the complex scalars $z^\alpha$ and
are determined by the prepotential $\mathcal{F}(z^\alpha)$ by the same formulae
as vector field couplings for vector multiplets. This reflects that the HM manifolds
resulting from CY3 compactifications are not generic, but of a special type, 
which can be obtained from a SK manifolds by the c-map. In the case at hand
the SK manifold is  the complex structure moduli space, and the c-map is either
the standard c-map or the temporal c-map. These c-maps can be defined
using the reduction of a Lorentz signature VM Lagrangian to three dimensions either
of space or over time. That the same types of HM manifolds occur in CY3
compactifications is no coincidence, but related to the fact that T-duality 
of type-II CY3 compactifications  exchanges VMs and HMs, as we will see later.
The HM geometry 
is QK for  type-IIA$_{(1,9)}$ and 
type-IIA$_{(0,10)}$ (and positive definite, since in our convention 
$\I_{IJ}$ is negative definite),
and PQK for type-IIA$^*_{(1,9)}$ and type-IIA$_{(2,8)}$. 
Note that in all cases the submanifold parametrized by the scalars $z^\alpha$
is an SK manifold.

\subsection{Type-IIB Calabi-Yau compactifications}

We now turn to type-IIB compactifications. 
The NS-NS sector is the same as for type-IIA. In the R-R sector 
the zero form $C_0$ gives rises to a scalar $c$. The two-form
$C_2$ gives rise to a two-form $C_{\mu \nu}$ which we dualize
into a scalar $a$, and to $h^{1,1}$ scalars $u^A$, $A=1, \ldots, h^{1,1}$. 
Taking into account the self-duality of the five-form $G_5$, 
the four-form $C_4$ gives rise to $h^{1,1}$ two-forms 
$C^A_{\mu\nu}$ which we dualize to 
scalars $v^A$, and $1 + h^{2,1}$ vectors $\mathcal{A}^0_\mu$
and $\mathcal{A}^\alpha_\mu$. The first vector is associated to 
the harmonic (3,0)-form of the CY3, while the other vectors correspond to the 
harmonic (2,1)-forms. See Table \ref{Tab:Massless_R-R-IIB} for a summary.

\begin{table}
\begin{center}
\begin{tabular}{| l | l| l| } \hline
   10 d & 4 d  & \\ \hline \hline 
   $C$ & c & scalar \\ \hline
   $C_{MN}$ & $C_{\mu \nu} \sim a$ & scalar \\
        & $C_{mn} \sim u^A$ &$h^{1,1}$ scalars \\ \hline 
   $C_{MNPQ}$ & $C_{\mu \nu mn} \sim v^A$ & $h^{1,1}$ scalars\\
                      & $C_{\mu mnp } \sim \mathcal{A}^0_\mu,
                        \mathcal{A}^A_{\mu}$ & $1+ h^{2,1}$ vector
                                               fields \\  \hline
 \end{tabular}
 \end{center}
 \caption{Massless fields in type-IIB Calabi-Yau compactifications, R-R sector.
   \label{Tab:Massless_R-R-IIB}}
\end{table}

Together with the NS-NS fields,
these fields are the bosonic content of the supergravity multiplet,
($g_{\mu \nu}, \mathcal{A}^0_\mu)$, of $h^{2,1}$ vector multiplets
($z^\alpha, A^\alpha_\mu$), and of $h^{1,1}+1$ hypermultiplets
($y^A, x^A, \varphi, \tilde{\phi}, u^A, v^A, c, a$). The vector fields
can be rearranged into linear combinations $A^I_\mu$ with
field strength $F^I_{\mu \nu}$ which together with the 
dual field strength $G_{I|\mu \nu}$ form a symplectic vector.
The relative signs between the kinetic terms are determined by 
those between the ten-dimensional fields and are listed in Table
\ref{Tab:IIBCY3}.

While one can perform the reduction of type-IIB theories explicitly, 
see for example \cite{Bohm:1999uk} for IIB$_{(1,9)}$, we can infer 
the result by using mirror symmetry and tracing sign flips. 
As is well known, type IIA$_{(1,9)}$ compactified on a CY3 with 
Hodge numbers $(h^{1,1}, h^{2,1})$ is equivalent to 
IIB$_{(1,9)}$ compactified on the mirror CY3 with Hodge numbers
$(h'^{1,1} = h^{2,1}, h'^{2,1}=h^{1,1})$. Both theories have 
$n_V = h^{1,1} = h'^{2,1}$ vector multiplets and $n_H=h^{2,1}+1 
= h'^{1,1} + 1$ hypermultiplets. In IIA compactifications 
complex structure module sit in hypermultiplets and K\"ahler moduli
in vector multiplets while in IIB compactifications it is the other
way round. We would like to compactify the IIB theory
on the same CY3 as the IIA theory, but this is the same as
compactifying the IIA theory on the mirror. The resulting theory 
has $n_V = h^{2,1}$ vector multiplets and $n_H = h^{1,1}+1$ 
hypermultiplets, and the action can be brought to our preferred standard form
of a vector and hypermultiplet action. 
To adapt results from type-IIB to type-IIB$^*$ and type-IIB', 
we then only have to trace the effect of the ten-dimensional 
sign flips. 

As a result, the bosonic Lagrangian for the supergravity multiplet and the $n_V = h_{2,1}$ 
vector multiplets takes the form
\begin{table}
\begin{center}
\begin{tabular}{| l | l || l | l | l | l | l | l | l | l || l |} \hline
 & $z^\alpha$ & $y^A$ & $x^A$ & $\varphi$ & $\tilde{\phi}$ & $c$ & $a$ & $u^A$ & $v_A$ & $A^I_\mu$ \\
 \hline
 IIB & + & + & + & + & + & + & + & + & + & + \\ \hline
 IIB$^*$ & + & + & + & + & + & $-$ & $-$ & $-$ & $-$ & $-$ \\ \hline
 IIB' & + & + & $-$  & + & $-$ & $-$ & +  & + & $-$ & $-$ \\ \hline
\end{tabular}
\end{center}
\caption{Signs of the kinetic terms for scalar and vector 
fields resulting from type-IIB CY3 compactifications. A  $+$ 
indicates a standard kinetic term. The fields $z^\alpha$ are the vector multiplet scalars.\label{Tab:IIBCY3}}
\end{table}

\begin{equation}
\label{4d_VM_Lagrangian_IIB}
\begin{aligned}
	L^{IIB/IIB^*/IIB'}_{G + VM} =   \half &\star R_4 - \bar{g}_{\alpha\bar{\beta}} (z, \bar{z}) dz^\alpha \wedge \star d \bar{z}^{\bar{\beta}} - \frac{\lambda}{4} \I_{IJ} F^I \wedge \star F^J
	+ \frac{1}{4} \cR_{IJ} F^I \wedge F^J\;,
\end{aligned}
\end{equation}	
where $\lambda=-1$ for IIB, and $\lambda=1$ for IIB$^*$ and IIB'. The geometry is
SK, with the two cases distinguished by an overall sign flip of the gauge fields.

In the hypermultiplet sector we can rearrange the scalars into linear 
combinations $\zeta^I \sim c, v^A$ and $\tilde{\zeta}_I \sim a, u^A$. 
The IIB HM Lagrangians take the form
\begin{equation}
\begin{aligned}
	L_{HM}^{IIB/IIB^*/IIB'} 
	=  &- \tilde{G}_{A \bar{B}} d z^A \wedge \star d \bar{z}^{\bar{B}} - \frac{1}{4} d \varphi \wedge \star d \varphi \\
	&+\epsilon_1  e^{-2 \varphi} 
		\left[d \tilde{\phi} + \half \left(\zeta^I d \tilde{\zeta}_I - \tilde{\zeta}_I d\zeta^I \right) \right] \wedge \star \left[d \tilde{\phi} + \half \left(\zeta^I d \tilde{\zeta}_I - \tilde{\zeta}_I d\zeta^I \right) \right] \\
		&-  \frac{\epsilon_2 }{2} e^{- \varphi} \left[ {\I}_{IJ} d \zeta^I \wedge \star d\zeta^J -
		\epsilon_1  {\I}^{IJ} \left(d \tilde{\zeta}_I + {\cR}_{IK} d \zeta^K \right)
		\wedge \star \left(d \tilde{\zeta}_I + {\cR}_{IK} d \zeta^K \right) \right] \;.
\end{aligned}
%\label{HM_master}
\end{equation}
For type-IIB the scalars $z^A=y^A + i x^A$ are complex and parametrize an SK submanifold. 
The  parameters $\epsilon_1, \epsilon_2$ take the values $(\epsilon_1,\epsilon_2)=(-1,-1)$, 
the HM manifold is positive definite and QK. 
For type-IIB$^*$ the signs of all R-R scalars are flipped, and $(\epsilon_1,\epsilon_2)=(-1,1)$.
The scalars $z^A$ are again complex and span an SK submanifold, and the HM manifold 
is PQK. When going from IIB$^*$ to IIB', the signs of $x^A,\tilde{\phi}$ and of $\tilde{\zeta}_I \sim a, u^A$ 
are flipped. The scalars $y^A$ and $x^A$ now combine into para-complex scalars
$z^A = y^A + e x^A$ which parametrize an SPK submanifold. We now have 
$(\epsilon_1,\epsilon_2)=(1,1)$. This is again a PQK manifold, but with a PSK submanifold
instead of an SK manifold. Thus the S-duality relating type-IIB$^*$ to type-IIB' changes
the HM manifolds in a significant way, while keeping it consistent with the same supersymmetry
algebra.

Comparing type-IIA with type-IIB compactifications on the same CY3, we see that, loosely speaking,
vector and hypermultiplet get exchanged. A type-IIA compactification has
$n_V=h^{1,1}$ vector and $n_H=h^{2,1}+1$ hypermultiplets, while a type-IIB 
compactification has $n'_V=h^{2,1}$ vector and $n'_H=h^{1,1}+1$ hypermultiplets.
As indicated by the Hodge numbers, complex structure moduli of the CY3 metric
end up in HMs for type-IIA and in VMs for type-IIB, while (para)-complexified
K\"ahler moduli end up in VMs for type-IIB and in HMs for type-IIA. In both cases
there is a universal HM which contains the dilaton, axion and two R-R fields. 
Moreover, in both cases all model dependence,
that is the dependence on the choice of the CY3, 
is encoded in two functions, the holomorphic prepotential of the complex 
structure moduli space and the (para)-holomorphic prepotential of the
(para)\-complexified K\"ahler moduli space.\footnote{To see that the two prepotentials
are on the same footing one must go beyond a simple dimensional reduction and 
include the $\alpha'$-corrections to the K\"ahler moduli space. We refer to \cite{Greene:1996cy}
for a review of string theory on Calabi-Yau manifolds.} The complex structure moduli space
is of course always complex, but the K\"ahler moduli space becomes para-complex
for type-IIA$_{(0,10)}$ and type-IIA$_{(2,8)}$ as well as type-IIB'$_{(1,9)}$ due to
the sign flip of the Kalb-Ramond field. The HM manifolds are completely determined
by their distinguished S(P)K submanifold through a c-map. This structure is consistent with 
certain pairs of compactifications being `on the same moduli' after compactification 
to three dimensions. As a result, type-II compactifications are mutually related 
by T-dualities transverse to the CY3. This will be studied in detail in the next section.

\section{T-duality}

We can now combine the results about c-maps with those about CY3
compactifications to determine how the four-dimensional theories
resulting from type-II CY3 compactifications are related by T-duality. 
Type-IIA$_{(1,9)}$  string theory compactified  on a circle of radius 
$R$, measured in string units $\sqrt{\alpha'}$, is equivalent to 
type-IIB$_{(1,9)}$ compactified on a circle of radius $1/R$ 
\cite{Dai:1989ua,Horava:1989ga}. Moreover, 
type-IIB$_{(1,9)}$ string theory on ten-dimensional Minkowski space 
can be obtained as an alternative decompactification limit $R\rightarrow 0$
of the circle compactified type-IIA$_{(1,9)}$ theory, with winding modes
playing the roles of momentum modes, and vice versa. This is what is meant when
saying that the uncompactified theories `are T-dual to each other.' T-duality extends
to backgrounds which include a compact factor transverse to the circle. 
In particular type-IIA$_{(1,9)}$ compactified on $X\times S^1_R$, is equivalent
to type-IIB$_{(1,9)}$ compactified on $X\times S^1_{1/R}$, where $X$ is the
same CY3. By taking
the alternative decompactification limit $R\rightarrow \infty$, one 
can map the four-dimensional effective field theories for 
 type-IIA$_{(1,9)}$  and  type-IIB$_{(1,9)}$, compactified on the same
 CY3  $X$, to one another, and the relation between the 
 respective vector and hypermultiplet sectors is given by the c-map and
 its inverse \cite{Cecotti:1988qn}. Timelike T-dualities and mixed 
 T-dualities which combine spacelike/timelike reduction with timelike/spacelike
 oxidation, together with S-duality, relate all ten-dimensional type-II theories to one another
 \cite{Hull:1998vg, Hull:1998ym}. In this section we extend these T-dualities to 
 CY3 compactifications. We remark that it is straightforward though somewhat
 tedious to work out the explicit relations between the fields of two T-dual
 four-dimensional effective field theories. T-duality operates naturally in the string frame, 
 and therefore we would need to convert our actions from the Einstein frame to the string frame,
 perform the reductions of T-dual theories over circles of radii $R$ and $1/R$, and then read 
 off the relations between the fields. 
 While the explicit map between fields is needed
 for some applications, in particular for mapping solutions from one theory to solutions of 
 a T-dual 
 theory, we will only be interested in how the various type-II CY3 compactifications are
 related to each other by T-duality and S-duality. For this it is sufficient to match 
 the hypermultiplet manifolds that we get after reduction to three dimensions, as this
show that both four-dimensional theories reduce to the same three-dimensional theory. 
 All that we need for this comparison was worked out in section 2. Explicit maps between 
 the fields will be given in a future publication where we will study the action of
 T-duality on solutions of the four-dimensional effective field theories.

\subsection{Signature (1,3) and spacelike/timelike T-duality}

To start exploring the web of relations between four-dimensional
theories we begin with the CY3 compactification of the  type-IIA$_{(1,9)}$  
theory.

\begin{figure}[!h]
	\centering
	\begin{tikzpicture}	
%		base
		\node (00) at (0,0) {A};
		\node (01) at (7,2) {A$^*$};
		\node (02) at (2,2) {B$^*$};
		\node (03) at (5,0) {B};
		\draw[semithick,<->] (00) -- (03) node [midway, below] {$C_S$};
		\draw[semithick,<->] (01) -- (02) node [midway, below] {$C_S$};
		\draw[semithick,<->] (01) -- (03) node [midway, right] {\; $C_T$};
		\draw[semithick,<->] (00) -- (02) node [midway, right] {\; $C_T$};
%		top
		\node (10) at (0,5) {IIA};
		\node (11) at (7,7) {IIA$^*$};
		\node (12) at (2,7) {IIB$^*$};
		\node (13) at (5,5) {IIB};
		\draw[semithick,<->] (10) -- (13) node [midway, above] {$T_S$};
		\draw[semithick,<->] (11) -- (12) node [midway, above] {$T_S$};
		\draw[semithick,<->] (11) -- (13) node [midway, right] {\; $T_T$};
		\draw[semithick,<->] (10) -- (12) node [midway, left] {$T_T$ \;};
%		legs
		\draw[semithick,<->] (00) -- (10) node [midway, left] {$CY3$};
		\draw[semithick,<->] (01) -- (11) node [midway, right] {$CY3$};
		\draw[semithick,<->] (02) -- (12) node [midway, left] {\; $CY3$};
		\draw[semithick,<->] (03) -- (13) node [midway, right] {$CY3$ \;};
	\end{tikzpicture}
	\caption{The spacelike and timelike T-dualities $T_S, T_T$ between the four type-II string theories in 
	ten-dimensional Minkwoski space induce relations between the four-dimensional supergravity theories, denoted A, B, A$^*$, B$^*$ obtained by compatification on the same Calabi-Yau threefold. 
The number $n_V$ of vector multiplets and $n_H$ of hypermultiplets is related to the Hodge number of the Calabi-Yau threefold by $(n_V, n_H) = (m,n) =(h_{1,1}, h_{2,1}+1) $ for type-A and $(n_V,n_H)=(m', n')=(h_{2,1}, h_{1,1}+1)$ for type-B. The theories denoted $A^*, \; B^*$ have the same structure, but
a modified supersymmetry algebra with 
a non-compact R-symmetry group which results in sign flips in the Lagrangian and modifications of the scalar geometry. 
The maps relating the four-dimensional theories are denoted $C_S$, $C_T$, depending on whether they  use a spacelike or timelike reduction and oxidation.  \label{Fig:cube}}
\end{figure}
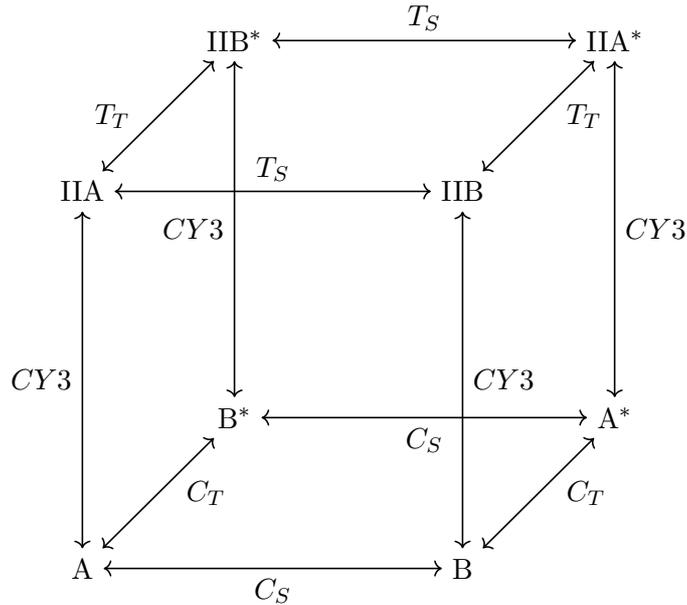

\begin{itemize}
\item
Type-IIA$_{(1,9)}$  string theory on a Calabi-Yau threefold
has $n_V = h^{1,1}$ vector and $n_H=h^{2,1}$ hypermultiplets. 
It realizes the 
standard $\mathcal{N}=2$ algebra with R-symmetry $U(2)\cong
U(1) \times SU(2)$ and the scalar manifold has the form
\[
\mathcal{M}^{IIA} = \mathcal{M}^{SK_+}_{2h^{1,1}} \times \tilde{\mathcal{N}}^{QK}_{4h^{2,1}+4} \;.
\]
Upon spacelike reduction the scalar manifold becomes
the product of two QK manifolds
\[
\mathcal{M}^{(1,2)} = \mathcal{N}^{QK}_{4h^{1,1} + 4} \times \tilde{\mathcal{N}}^{QK}_{4h^{2,1}+4} \;.
\]
If one swaps the roles of the two factors and lifts back over space, one obtains
\[
\mathcal{M}^{IIB} = \mathcal{N}^{QK}_{4h^{1,1}+4 } \times \tilde{\mathcal{M}}^{SK_+}_{2 h^{2,1}}
\]
as required for 
a Calabi-Yau compactification of type-IIB string theory. 
This is the standard, spatial T-duality between type-IIA and type-IIB, extended
to their Calabi-Yau compactifications.  It employs the standard, spatial c-map 
in both directions. 
\item
If we start again with type-IIA, but perform a timelike reduction, we obtain a theory
in signature (0,3) with scalar target 
\[
\mathcal{M}^{(0,3)} = \mathcal{N}^{PQK}_{4h^{1,1} + 4} \times \tilde{\mathcal{N}}^{QK}_{4h^{2,1}+4}\;,
\]
where the first factor is now PQK rather than QK. Swapping the two 
factors and lifting back over time we obtain a scalar manifold of the form
\[
\mathcal{M}^{IIB^*} = \mathcal{N}^{PQK}_{4h^{1,1}+4} \times \tilde{\mathcal{M}}^{SK_-}_{2h^{2,1}}  \;.
\]
Note that after the oxidation we have flipped gauge field terms (recorded as SK$_-$)
since we need such a sign
in order to obtain a QK manifold by timelike reduction. The resulting four-dimensional
theory realizes the twisted Lorentz signature algebra with R-symmetry $U(1,1) \cong
U(1) \times SU(1,1)$. Thus we obtain the timelike T-duality between type-IIA and type-IIB$^*$, 
extended
to their Calabi-Yau compactifications. It employs the temporal c-map for reduction and the
twisted temporal c-map for oxidation. 
\item
If we start with type-IIA$^*$ the initial scalar manifold is
\[
\mathcal{M}^{IIA^*} = \mathcal{M}^{SK_-}_{2h^{1,1}} \times \tilde{\mathcal{N}}^{PQK_{SK}}_{4h^{2,1}+4} \;.
\]
Upon space-like reduction this becomes
\[
\mathcal{M}^{(1,2)} = \mathcal{N}^{PQK_{SK}}_{4h^{1,1} + 4} \times \tilde{\mathcal{N}}^{PQK_{SK}}_{4h^{2,1}+4} \;,
\]
which lifts back to
\[
\mathcal{M}^{IIB^*} = \mathcal{N}^{PQK_{SK}}_{4h^{1,1}+4} 
\times \tilde{\mathcal{M}}^{SK_-}_{2h^{2,1}}  \;.
\]
This realizes the spacelike T-duality between type-IIA$^*$ and type-IIB$^*$,
extended to their Calabi-Yau compactifications. Here we employ the twisted spatial 
c-map in both directions. 
\item
If we start with type-IIA$^*$ and reduce over time we obtain instead
\[
\mathcal{M}^{(0,3)} = \mathcal{N}^{QK}_{4h^{1,1}+ 4} \times 
\tilde{\mathcal{N}}^{PQK_{SK}}_{4h^{2,1}+4} \;,
\]
which lifts back to 
\[
\mathcal{M}^{IIB} = \mathcal{N}^{QK}_{4h^{1,1} + 4} \times \tilde{\mathcal{M}}^{SK_+}_{2h^{2,1}} \;,
\]
and we realize the timelike T-duality between Calabi-Yau compactifications 
of type IIA$^*$ and type IIB. Here we use the twisted temporal c-map for reduction 
and the temporal c-map for oxidation.
\end{itemize}
The relations between the four-dimensional theories are summarized by the
lower face of the cubic diagram in Figure
\ref{Fig:cube}.

\subsection{Mixed T-dualities and signature change}

Let us now mix spacelike/timelike reduction with timelike/spacelike oxidation
in order to relate four-dimensional theories across signatures. 
If we start with the CY3 compactification of type-IIA$_{(0,10)}$ we have
a scalar manifold which is the product of an SPK and a QK manifold:
\[
\mathcal{M}^{IIA, (0,4)} = \mathcal{M}^{SPK}_{2h^{1,1}} \times \tilde{\mathcal{N}}^{QK}_{4h^{2,1}+4} \;.
\]
Upon spacelike reduction to signature $(0,3)$, 
the scalar manifold becomes
\[
\mathcal{M}^{(0,3)} = \mathcal{N}^{PQK_{SPK}}_{4h^{1,1}+4} \times 
\tilde{\mathcal{N}}^{QK}_{4h^{2,1}+4} \;,
\]
where the first  PQK manifold has an SPK base. This step involves the Euclidean c-map. 
We now need to identify a IIB theory that gives
rise to the same scalar manifold upon timelike reduction. To obtain the QK 
manifold $\tilde{\mathcal{N}}^{QK}_{4h^{2,1}+4}$ by timelike reduction we need to 
start with vector multiplets which have SK geometry and a sign flip
between scalar and vector term, denoted SK$_-$.  This could be either the
CY3 compactification of IIB$^*$ or IIB'. The map 
$\tilde{\mathcal{M}}^{SK_-}_{2h^{2,1}} \rightarrow \tilde{\mathcal{N}}^{QK}_{4h^{2,1}+4} $
is the twisted version of the temporal c-map. The HM manifold of 
the partner theory must match $\mathcal{N}^{PQK_{SPK}}_{4h^{1,1}+4}$,
that is, it must be a PQK manifold with an SPK base. Therefore
we need to choose IIB', which has a scalar manifold of the type
\[
\mathcal{M}^{IIB'} = \mathcal{N}^{PQK_{SPK}}_{4h^{2,1}+4} \times \tilde{\mathcal{M}}^{SK_-}_{2h^{1,1}} \;.
\]
This shows the existence of a mixed T-duality relating the CY3 compactifications 
of type-IIA$_{(0,10)}$ and type-IIB'$_{(1,9)}$, which uses the Euclidean c-map for
reduction and the twisted temporal c-map for oxidation.

If we reduce the IIB' theory over space, the resulting scalar manifold is
\[
\mathcal{M}^{(1,2)} = \mathcal{N}^{PQK_{SPK}}_{4 h^{2,1}} 
\times \tilde{\mathcal{N}}^{PQK_{SK}}_{4h^{1,1}+4} \;,
\]
where $\tilde{\mathcal{M}}^{SK_-}_{2h^{1,1}} \rightarrow \tilde{\mathcal{N}}^{PQK_{SK}}_{4h^{1,1}+4}$ 
is the twisted version of the spatial c-map. 

By lifting this back over time we obtain the
 scalar manifold of  the CY3 compactification of IIA$_{(2,8)}$,
\[
\mathcal{M}^{IIA, (2,2)}  = \mathcal{M}^{SPK}_{2h^{1,1}} \times 
\tilde{\mathcal{N}}^{PQK_{SK}}_{4h^{2,1}+4} \;.
\]
This step involves the inverse of the neutral c-map which maps SPK to PQK through
a timelike reduction. In summary we have shown the existence of a mixed T-duality 
relating the CY3 compactifications of type-IIB'$_{(1,9)}$ and type-IIA$_{(2,8)}$ 
which uses the twisted spatial c-map for reduction and the neutral c-map for 
oxidation. We summarize the six T-dualities which relate type-IIA and type-IIB theories
in four dimensions in Table \ref{Tab:T-dualities}. Note that under T-duality 
the compactifications organise into two orbits: the orbit of `pure' T-dualities relating
IIA/IIA$^*$/IIB/IIB$^*$ in signature (1,3), and the orbit of mixed T-dualities
relating type-IIA theories in signatures (0,4) and (2,2) to the type-IIB' theory
in signatures (1,3). In order connect these two orbits to one another we would
need to use the duality between IIB$^*$ and IIB', which is an S-duality.
Since CY3 backgrounds only preserve four-dimensional 
$\mathcal{N}=2$ supersymmetry, there is not good reason to expect 
that S-duality is valid, and therefore we should expect that there are 
two distinct classes of compactifications. The relation of the backgrounds
within each orbit relies on T-duality for backgrounds of the form 
CY3 $\times$ $S^1$ which is an established
perturbative symmetry of string theory. Note however, that there are special, 
non-generic $\mathcal{N}=2$ compactifications which are `$\mathcal{N}=4$-like'
and exhibit S-duality. For this class all type-II CY3 compactifications should
form a single orbit at the non-perturbative level by combining  pure T-dualities,
mixed T-dualities  and S-duality. Note that for S-duality to work, there
has to exist, for the HM manifolds of these special models,
an isomorphism of PQK manifolds, which replaces an SK base with 
an SPK base. Such an isomorphism can only exist in special case when
the structure of the prepotential for the S(P)K base is very simple.

\begin{table}
{\small
\begin{tabular}{| l | l || l | l || l | l |} \hline 
IIA in 4d & Scalar mfd. & 3d sig. & Scalar mfd. & IIB in 4d & Scalar mfd.  \\ \hline \hline
IIA$_{(0,4)}$ & SPK $\times$ QK & (0,3) & PQK$_{SPK}$ $\times$ QK 
& IIB'$_{(1,3)}$ & PQK$_{SPK}$ $\times$ SK$_-$ \\ \hline
IIA$_{(1,3)}$ & SK$_+$ $\times$ QK & (0,3) & PQK$_{SK}$ $\times$ QK 
& IIB$^*_{(1,3)}$ & PQK$_{SK}$ $\times$ SK$_-$ \\ \hline
IIA$^*_{(1,3)}$ & SK$_-$ $\times$ PQK$_{SK}$  & (0,3) & QK $\times$ PQK$_{SK}$ 
& IIB$_{(1,3)}$ & QK  $\times$ SK$_+$ \\ \hline
IIA$_{(1,3)}$ & SK$_+$ $\times$ QK  & (1,2) & QK $\times$ QK 
& IIB$_{(1,3)}$ & QK  $\times$ SK$_+$ \\ \hline
IIA$^*_{(1,3)}$ & SK$_-$ $\times$ PQK$_{SK}$  & (1,2) & PQK$_{SK}$ $\times$ PQK$_{SK}$ 
& IIB$^*_{(1,3)}$ & PQK$_{SK}$  $\times$ SK$_-$ \\ \hline
IIA$_{(2,2)}$ & SPK $\times$ PQK$_{SK}$  & (1,2) & PQK$_{SPK}$ $\times$ PQK$_{SK}$ 
& IIB'$_{(1,3)}$ & PQK$_{SPK}$  $\times$ SK$_-$ \\ \hline
\end{tabular}
\caption{Summary of T-dualities between type-IIA and type-IIB Calabi-Yau compactifications
for all (inequivalent) four-dimensional signatures. In the middle we specify the 
three-dimensional theory to which the T-dual four-dimensional theories reduce. \label{Tab:T-dualities}}
}
\end{table}

Let us finally point out that considering the other ten-dimensional signatures
will add nothing new.  The type-II$_{(5,5)}$ theories cannot be compactified
on CY3 folds, and the theories IIA$_{(10,0)}$, II$_{(9,1)}$, IIA$_{(8,2)}$
are related to those we have considered by an overall sign change of the
metric, which maps  $(t,s) \rightarrow (s,t)$. Theories related in this way
have been shown  to be equivalent \cite{Hull:1998ym}. The
type-IIA$_{(4,6)}$ and type-IIA$_{(6,4)}$ reduce to theories in signature (4,0) and 
(0,4) which are equivalent to those we have considered. From the higher-dimensional 
point of view the chain of mixed T-dualities is projected onto the chain we have
described. Since the four-dimensional theory in signature (0,4) is unique,
the difference between type-IIA$_{(0,10)}$ and type-IIA$_{(4,6)}$ 
is lost from the four-dimensional perspective. 
Similarly the  type-IIB$_{(3,7)}$ and type-IIB$_{(7,3)}$ reduce to theories in 
signatures $(3,1)$ of (1,3), which are of the type-IIB' type, and the 
distinction between type-IIB'$_{(3,7)}$  and  type-IIB'$_{(1,9)}$ is lost from 
the four-dimensional perspective.\footnote{The actions of IIB'$_{(1,9)/(9,1)}$  
and IIB'$_{(3,7)/(7,3)}$ differ by an overall sign flip of the R-R fields \cite{Hull:1998ym,Dijkgraaf:2016lym}.
Upon compactification on a CY3, this results in a flip of the parameter $\epsilon_2$ in the
hypermultiplet sector, but the resulting hypermultiplet manifolds are isometric.} 
We have already pointed out
that in order to relate such a theory to type-IIB$^*$, and thus to the other T-duality 
orbit, we need to use S-duality, which can only be expected to be a symmetry for
non-generic $\mathcal{N}=2$ compactifications. Signatures (3,7) and (7,3)
have a unique theory of type-IIB', but they can be related through mixed T-dualities
to signatures (1,9) and (9,1) where by using S-duality they can be related
to type-IIB$^*$ and from there by T-dualities to type IIA/IIA$^*$ and type-IIB. 
Thus for $\mathcal{N}=2$
compactifications which preserve S-duality, we can connect the CY3 compactifications
of all type-II theories in signatures (0,10) , \ldots (4,6) to one another (and the same
applies to signature obtained by an overall sign flip). For generic $\mathcal{N}=2$ we have to expect
two disjoint T-duality orbits.

\section{Outlook}

In this paper we have obtained the CY3 compactifications of all 
ten-dimensional type-II string theories and analyzed how they are related
to one another by T-duality and S-duality. At the level of symmetries and 
effective supergravity, we get a full and satisfactory picture which 
is consistent with the idea that exotic string theories and their
compactifications fit into an extended string theory landscape and
allow one to realize all maximally supergravity theories in all signatures
as limits. Of course, admitting backgrounds with multiple time directions,
as well as inverted kinetic terms in the effective action raises conceptual
questions. Instead of repeating the arguments of \cite{Hull:1998vg,Hull:1998ym,Dijkgraaf:2016lym,Blumenhagen:2020xpq},
let us ask what new insights and future directions result from our work. 

Apart from symmetry considerations, a reason to consider the 
inclusion of exotic string theories into the string theory landscape 
is string universality, and the observation that bubbles of
exotic spacetime signature can be generated \cite{Dijkgraaf:2016lym}. One future
direction is to explore in more detail whether and how dynamical signature change can 
be realized as a physical process in string theory.  It would be interesting to relate
this to the recent work \cite{Kontsevich:2021dmb,Witten:2021nzp,Lehners:2021mah}
on complex spacetime metrics, which was in part motivated by earlier
work \cite{Louko:1995jw} on topology change. Gravity with a dynamical signature
has recently been discussed in \cite{Bondarenko:2021xvz} within the framework
of Einstein-Cartan gravity.

To make a preliminary remark on this topic, we note that 
the T-duality orbit along which 
spacetime signature can change in type-II CY3 compactifications is connected to the orbit of 
the standard IIA/IIB compactifications and their IIA$^*$/IIB$^*$ partners
by S-duality, which we cannot expect to be valid in generic $\mathcal{N}=2$
compactifications. This suggests that in backgrounds with less than
$\mathcal{N}=4$ supersymmetry, there is, generically (with the 
exception of special `$\mathcal{N}=4$ like' backgrounds), a separation 
between a phase with Lorentzian string worldsheets and fixed Lorentzian
spacetime signature, and a phase with Euclidean string worldsheets and 
arbitrary spacetime signature.\footnote{Lorentizan string worldsheets are 
also possible in neutral signature (5,5), but this signature does not give
rise to CY3 compactifications. Also, to connect it to the standard IIA/IIB
theories, one needs to use S-duality.} 
Thus signature change may only be 
relevant cosmologically if the universe goes through a phase of high
unbroken supersymmetry.\footnote{Our universe could still be a 
brane world embedded into a higher-dimensional universe with 
multiple time directions, an option that has been explored in  \cite{Blumenhagen:2020xpq}.} 

Based on the results of this paper, 
 solutions of exotic string theories can now be explored systematically 
from a four-dimensional $\mathcal{N}=2$ perspective 
in addition to the ten-dimensional perspective. Future work will
include how solutions transform under dualities, whether solutions
of different theories can be connected to one another, including the
question of dynamical signature change. Staying within the class
of theories with Lorentzian string worldsheets, there are interesting
questions regarding the relation between solutions in type-II and
type-II$^*$. The results of our paper imply that the dual pair
of planar cosmological and black hole solutions described in 
\cite{Gutowski:2019iyo,Gutowski:2020fzb} lifts to a `dual pair' 
of solutions in type-IIA and type-IIA$^*$. This raises the question 
whether these `dual solutions' can be related by T-duality (which is not
obvious). Both solutions have horizon thermodynamics that can
be related to the same Euclidean thermal partition functions, and
one can now ask whether the type-II embedding provides insights
into the underlying microscopic physics. One could also study
whether these solutions correspond to admissible saddle points, 
in the sense of \cite{Kontsevich:2021dmb,Witten:2021nzp,Lehners:2021mah},
of the Euclidean path integral.

\subsubsection*{Acknowledgements}
T.M. thanks Owen Vaughan for stimulating discussions at an early 
stage of this project.

\appendix

%%%%%%%%%%%%%%%%%%%%%%%%%%%%%%%%%%%%%%%%%

\section{Calabi-Yau compactifications}

In this appendix we provide some more details on Calabi-Yau compactifications 
of type-IIA theories. Since this has been worked out in detail for IIA$_{(1,9)}$ 
in \cite{Bodner:1990zm} and for IIA$_{(0,10)}$ in \cite{Sabra:2015tsa}  
all we need in order to include the additional cases of IIA$^*_{(1,9)}$ and 
IIA$_{(2,8)}$, is to trace the ten-dimensional sign flips through
the computation. The results from the reduction of the individual terms
is taken from \cite{Sabra:2015tsa} whose conventions and notation we follow.
We refer to \cite{Huebsch:1991,Candelas:1990pi,LopesCardoso:2019mlj} for 
further background on CY3 compactifications and special geometry.

\subsection{Calabi-Yau threefolds} 
A Calabi-Yau threefold (CY3) $X$ is a compact complex manifold of (complex) dimension 
three which admits a Ricci-flat K\"ahler metric. This condition is equivalent
to the existence of a nowhere vanishing holomorphic (3,0)-form $\Omega$. 
We denote local real coordinates on $X$ by $y^a$, $a=1, \ldots, 6$ and
introduce complex coordinates $\xi^i$, $i=1,2,3$ 
\begin{equation*}
	\xi^1 = \frac{1}{\sqrt{2}} (y^1 + i y^2), \quad \xi^2 = \frac{1}{\sqrt{2}} (y^3 + i y^4), \quad \xi^3 = \frac{1}{\sqrt{2}} (y^5 + i y^6).
\end{equation*}
The components of the K\"ahler metric in complex coordinates are $g_{i\bar{j}}$, 
and the volume form is
\[
\mbox{vol}_g = \sqrt{g} \, d^6y = i \, \sqrt{g} \, d^3\xi d^3 \bar{\xi}  \;.
\]
Therefore the volume of $X$ is
\[
\mathcal{V} =
\int_X \mbox{vol}_g = \int_X \sqrt{g} \, d^6y = \int_X i \, \sqrt{g} \, d^3\xi d^3 \bar{\xi}  \;.
\]
The scalar product between differential $(p,q)$-form is 
\begin{equation*}
	(\omega_{(p,q)}, \eta_{(p,q)}) = \int_{X} \omega_{(p,q)} \wedge \star \, \eta_{(p,q)} \;,
\end{equation*}
where $\star$ is the Hodge operator with respect to the K\"ahler metric $g_{i\bar{j}}$.
For reference we note that
\begin{equation*}
	\star \rho_{(3,0)} = -i \rho_{(3,0)}, \qquad \sigma_{(2,1)} = i \sigma_{(2,1)} \;.
\end{equation*}
The massless spectrum of a CY3 compactification is determined by the harmonic
$(p,q)$-forms. On a compact K\"ahler manifold, harmonic $(p,q)$-forms represent
elements of the Dolbeault cohomology groups $H^{p,q}_{\bar{\partial}}(X, \mathbb{C})$. 
The Hodge numbers $h^{p,q} = \dim H^{p,q}_{\bar{\partial}}(X, \mathbb{C})$ of a
CY3 are
\begin{eqnarray*}
&& h^{0,0} = h^{3,0} = h^{0,3} = h^{3,3} = 1 \;,\;\;\;
h^{1,0}=h^{0,1} = h^{3,2}=h^{2,3} =0 \;,\;\;\\
&& h^{1,1}=h^{2,2}\geq 1, \;\;h^{1,2}=h^{2,1}\geq 0\;.
\end{eqnarray*}
We introduce a basis for harmonic forms representing Dolbeault cohomology classes:
\begin{equation*}
\begin{aligned}
	&(1,1) \qquad V^A = V^A_{i\bar{j}} d \xi^i \wedge d\bar{\xi}^j, \qquad &&A = 1,\ldots, h_{1,1}\\
	&(2,1) \qquad \Phi_\alpha = \frac{1}{2} \Phi_{\alpha ij\bar{k}} d \xi^i \wedge d \xi^j \wedge d\bar{\xi}^k, \qquad \qquad &&\alpha = 1,\ldots, h_{2,1}\\
	&(3,0) \qquad \Omega = \frac{1}{3!} \Omega_{ijk} d \xi^i \wedge d \xi^j \wedge d \xi^k \;.\\
\end{aligned}
\end{equation*}
For the harmonic three-forms we also introduce a real basis $\alpha_I, \beta^J$, 
where $I = 0,\ldots,h_{2,1}$, which is dual to a canonical basis of 
third homology group $H_3(X,\mathbb{Z})$:
\begin{equation*}
\int_{A^I} \alpha_J = 	\int_{X} \alpha_I \wedge \beta^J = \delta^J_I =  A_I \cdot B^J \;\;\;
\int_{B_I} \beta^J = \int_{X} \beta^J \wedge \alpha_I = - \delta^J_I  = - B^J \cdot A_I \;.
\end{equation*}
Here `$\cdot$' denotes the intersection product of three-cycles.
Observe that $H_3(X,\mathbb{Z})$ and $H^3(X,\mathbb{R})$ 
 carry a natural symplectic structure.

Since $X$ is a K\"ahler manifold, once we have fixed a  complex structure the metric
is determined by the choice of a K\"ahler form. This is a harmonic (1,1)-form
$J = M^A V^A$ which satisfies the positivity requirements
\begin{equation}
\mathcal{K} = \int_X J \wedge J \wedge J = 6 \mathcal{V} > 0  \;, \;\;\;
 \int_{D^A} J \wedge J = \int_X V_A \wedge J \wedge J >0  \;,\;\;\;
\int_{C_A} J = \int V^A \wedge J > 0 \;,
\end{equation}
where $C_A$ and $D^A$ are generators of $H_2(X, \mathbb{Z})$ and 
$H_4(X,\mathbb{Z})$ and $V_A$, $V^A$, ar the dual generators
of $H^4(X,\mathbb{R})$ and $H^2(X,\mathbb{R})$. Thus the positivity
conditions ensure that the volumes of $X$ as well as those of all surfaces
and curves in $X$ are positive. The K\"ahler form is related to the metric
by 
\[
J = i g_{i\bar{j}} d\xi^i \wedge d\bar{\xi}^{\bar{j}} \;.
\]

Since the metric $g_{i\bar{j}}$ is K\"ahler, its deformations split up into two types. 
\begin{enumerate}
\item
Deformations of the form $\delta g_{i\bar{j}}$ preserve the complex structure. 
When imposing in addition that Ricci flatness is preserved, the independent deformations
(after taking into account reparametrizations) can be parametrized as
\[
i \delta g_{i\bar{j}} = \sum_{A=1}^{h^{1,1}} \delta M^A V^A \;,
\]
where $M^A$ are real parameters and $V^A$ is our basis for the harmonic
(1,1)-forms. Thus these deformations correspond to deformations of the 
K\"ahler form. 
\item
Deformations of the form $\delta g_{ij}, \delta g_{\bar{i} \bar{j}} = \overline{\delta g_{ij}}$ 
change the complex structure. 
Inequivalent changes of the complex structure are parametrized 
by elements of $H^1(X,TX) \cong H^{2,1}(X)$, where we used that
the holomorphic top form can be used to identify vector-valued (0,1)-forms
with (2,1) forms:
\[
\phi_{ij\bar{k}} := \Omega_{ijl} \psi^l_{\bar{k}}
\]
The inequivalent complex structure changing  deformations of
the metric can be parametrized using harmonic (2,1)-forms:
\[
\delta g_{\bar{i}\bar{j}} = \sum_{\alpha=1}^{h^{2,1}} \delta z^\alpha 
b_{\alpha \bar{i}\bar{j}} \;,\;\;
\mbox{where}\;\;\;
b_{\alpha} = \frac{1}{2} b_{\alpha \bar{i}\bar{j}} d \bar{\xi}^{\bar{i}} \wedge d\bar{\xi}^{\bar{j}}
= -\frac{i}{2} \frac{1}{ ||\Omega||^2} \bar{\Omega}_{\bar{i}}^{\;\;kl} \Phi_{\alpha kl \bar{j}}
d \bar{\xi}^{\bar{i}} \wedge d\bar{\xi}^{\bar{j}} \;,
\]
where $z^\alpha$ are complex parameters, where
\[
||\Omega||^2 = \frac{1}{3!} \Omega_{ijk} \bar{\Omega}^{ijk} 
\]
and where $\Phi_\alpha$ is our basis for the harmonic (2,1)-forms. 
\end{enumerate}
In CY3-compactifications the parameters $M^A, z^\alpha$ become scalar fields,
which appear in the action through sigma models whose target spaces are 
the moduli spaces of K\"ahler structures and of complex structures, equipped 
with their natural metrics, about which we will report next. The so-called Weil-Peterson 
metric on the space of complex structures can be expressed through scalar products
of harmonic forms as follows:
\[
g_{\alpha \bar{\beta}} = \frac{1}{{\cal V}}  (b_\alpha, \bar{b}_{\bar{\beta}}) =
\frac{1}{||\Omega||^2 {\cal V}} (\Phi_\alpha, \bar{\Phi}_{\bar{\beta}}) =
\frac{ - i \int_X \Phi_\alpha \wedge \bar{\Phi}_{\bar{\beta}}}{i \int_X \Omega \wedge
\bar{\Omega} } \;.
\]
The metric $g_{\alpha \bar{\beta}}$ is a K\"ahler metric, in fact a projective special
K\"ahler metric of precisely the same type as appears for the target spaces of
four-dimensional vector multiplets coupled to supergravity. 
To make this structure explicit, one notes that 
the complex structure of $X$ is encoded by the `direction' of $H^{3,0}_{\bar{\partial}}(X,\mathbb{C})$
inside $H^3(X,\mathbb{C})$. To parametrize the choice of a complex structure one can use
the periods of the holomorphic top-form $\Omega \in  H^{3,0}_{\bar{\partial}}(X,\mathbb{C})
\subset H^3(X,\mathbb{C})$:
\[
X^I = \int_{A_I} \Omega \;,\;\;\;
F_J = \int_{B^J} \Omega \;.
\]
The top form can be expanded as
\begin{equation*}
	\Omega = X^I \alpha_I - F_I \beta^I \;.
\end{equation*}

The periods $X^I, F_J$ are not independent. Without loss of generality, 
we can assume that the  periods $X^I$ are independent and parametrize the 
inequivalent choices of a holomorphic top-form $\Omega$ among the three-forms.
Then the  periods $F_J$ are, at least locally,  functions of the coordinates $X^I$, and 
for a generic choice of a basis, they form the gradient $F_J = \partial{F}/\partial X^J$ of
a function $F(X^I)$, the prepotential,  
which is holomorphic and homogeneous of degree two.\footnote{This generic situation 
can always be obtained by applying a symplectic transformation.}
 Holomorphic
top forms which differ by a complex scalar factor, $\Omega \rightarrow \lambda \Omega$,
$\lambda \in \mathbb{C}^*$ define the same complex structure. Therefore the $X^I$ 
are projective coordinates on the space $\mathcal{M}_{2h^{2,1}}$ of complex
structures. So-called special coordinates on $\mathcal{M}_{2h^{2,1}}$, which 
parametrize inequivalent complex structures, are provided
by the ratios $z^\alpha = X^\alpha/X^0$. The space parametrized by the periods
$X^I$ is the space of holomorphic top-forms, which forms, at least locally, 
a complex cone $\mathcal{CM}_{2h^{2,1}+2}$ over the moduli space of 
complex structures. Here we see the same type of special geometry arising which also 
characterizes the scalar geometry of four-dimensional vector multiplets. 
$\mathcal{CM}_{2h^{2,1}+2}$ is a conical affine special K\"ahler (CASK) manifold,
while $\mathcal{M}_{2h^{2,1}}$ is the associated projective special K\"ahler
manifolds. The special K\"ahler metrics on both spaces can be expressed
using the prepotential $F(X^I) = (X^0)^2 \mathcal{F}(z^A)$. The associated
K\"ahler potentials $K_{CASK}, K$ and metrics $N_{IJ}, g_{\alpha \beta}$ are
\begin{eqnarray}
K_{CASK}  =  i (X^I \bar{F}_I - F_I \bar{X}^I) 
&\Rightarrow& N_{IJ} = \frac{\partial^2 K_{CASK}}{\partial X^I \partial \overline{X^J}} = 
2 \mbox{Im} F_{IJ} \;, \\
K  = - \log
\left( -i (X^I \bar{F}_I - F_I \bar{X}^I ) \right)  &\Rightarrow& g_{\alpha \bar{\beta}} 
= \frac{\partial^2 K}{\partial z^\alpha \partial \bar{z}^{\bar{\beta}}}  \;.
\end{eqnarray}
Note that since $F(X^I)$ is homogeneous of degree two, its derivatives $F_I, F_{IJ}, \ldots$
are homogeneous of degrees one, zero, \ldots. This allows one to rewrite formulas
in various ways, for example $K = -\log(\bar{X}^I N_{IJ} X^J)$. 

These are generic expressions for special K\"ahler geometry, which are 
valid without any reference to CY3-folds.  In CY3 compactifications
the special geometry data can be expressed in terms of CY3 data. In particular:
\[
(\Omega, \bar{\Omega}) = i \int_X  \Omega \wedge \bar{\Omega} = 
||\Omega||^2 {\cal V} = - i (X^I \bar{F}_i - F_I \bar{X}^I) = 
-N_{IJ} \bar{X}^I X^J \;,
\]
where $N_{IJ}$ is the CASK metric on $\mathcal{CM}_{2h^{1,2}+2}$. 
The K\"ahler potential can be expressed in terms of the top form:
\[
K = - \log ( \Omega, \bar{\Omega}) = - \log \left( i \int_X \Omega \wedge \bar{\Omega} \right)
\Rightarrow g_{\alpha \bar{\beta}} = \frac{\partial^2}{\partial z^\alpha \partial \bar{z}^{\bar{\beta}}}
\left( - \log \left( i  \int_X \Omega \wedge \bar{\Omega} \right)\right) \;.
\]

Let us now consider deformations of the K\"ahler form.  Following \cite{Bodner:1990zm}, we define the following integrals 
\begin{equation*}
\begin{aligned}
		\K &= \int_{CY_3} J \wedge J \wedge J\;, & \K_A &= \int_{CY_3} V^A \wedge J \wedge J  \;,\\
		\K_{AB} &= \int_{CY_3} V^A \wedge V^B \wedge J\;, \qquad & \qquad \K_{ABC} &= \int_{CY_3} V^A \wedge V^B \wedge V^C \;.\\
\end{aligned}
\end{equation*}
We remark that $\K$ is proportional the volume $\V$, $\mathcal{K} = 3! \mathcal{V}$, while 
$\K_{ABC}$ are the triple intersection numbers of homology four-cycles.  
The Hodge dual of a $(1,1)$ has the form \cite{Candelas:1990pi}
\begin{equation*}
	\star V^B = - J \wedge B^B + \frac{3}{2 \K} J \wedge J \left( \int_{CY_3} V^B \wedge J \wedge J \right).
\end{equation*}
This allows us to evaluate the inner product of two $(1,1)$ forms:
\begin{equation*}
	G_{AB} (M) := \frac{1}{2 \V} \int_{CY_3} V^A \wedge \star V^B = -3 \left( \frac{\K_{AB}}{\K} - \frac{3}{2} \frac{\K_A \K_B}{\K^2} \right) \;.
\end{equation*}
Using the expansion $J = M^A V^A$ of the K\"ahler form, we obtain
\begin{eqnarray*}
&& \K = \K_{ABC} M^A M^B M^C =: (\K MMM)\;,\;\;\;
\K_A = \K_{ABC} M^B M^C  =: (\K MM)_A \;,\;\;\; \\
&& \K_{AB} = \K_{ABC} M^C  =: (\K M)_{AB} \;,
\end{eqnarray*}
so that
\begin{equation}
G_{AB}(M) = - 3 \left(
\frac{ (\K M)_{AB}}{ (\K MMM)} - \frac{3}{2} \frac{ (\K MM)_A (\K MM)_B}{(\K MMM)^2} \right) \;. 
\end{equation}
This shows that the metric $G_{AB}(M)$ is a conical affine special real metric with Hesse potential 
$\mathcal{K}$, which implies that its  pullbacks to hypersurfaces $\mathcal{K}=$ const. 
are projective special real metrics. 
These are the general target spaces of five-dimensional 
vector multiplets, which reflects that spaces of K\"ahler moduli with overall volumes
fixed appear as the vector multiplet moduli spaces of eleven-dimensional supergravity
compactified to five dimensions on a CY3 \cite{Cadavid:1995bk}. 
We refer to \cite{LopesCardoso:2019mlj}  for more details on special real geometry.

In type-II CY3 compactifications the K\"ahler moduli space, which naturally is a 
real space with a Hessian metric, is extended to a special K\"ahler
or a special para-K\"ahler space once the $B$-field moduli are taken into account.
In theories of closed oriented strings, the metric $G_{MN}$ is accompanied by the 
Kalb-Ramond field $B_{MN}$. For CY3 compactifications the $B$-field contributes
$h^{1,1}$ real scalars associated to harmonic (1,1)-forms, which combine with 
the $h^{1,1}$ K\"ahler moduli $M^A$. If the $B$-field has a standard kinetic
term, as is the case for type-IIA/IIA$^*$/IIB/IIB$^*$ in signature (1,9), 
then including the $B$-field leads to a complexification of the K\"ahler
form and of the K\"ahler moduli space. The resulting moduli space is 
projective special K\"ahler, and completely determined by the underlying 
real K\"ahler moduli space by the supergravity r-map, which maps
projective special real manifolds to projective special K\"ahler manifolds. 
If the the sign of the kinetic term of the $B$-field is flipped, as it happens
for type-IIA$_{(0,10)}$, type-IIB' and type-IIA$_{(2,8)}$, this induces a
sign flip for the $h^{1,1}$ $B$-field moduli. As a result, the K\"ahler moduli
space is para-complexified rather than complexified. The resulting 
projective special para-K\"ahler manifold is obtained from the real K\"ahler
moduli space by the para-r-map, which maps projective special real 
manifolds to special para-K\"ahler manifolds. 

Thus by combining the scalar fields descending from the metric and the $B$-field,
we end up either with two SK manifolds or with one SK and one SPK manifold, 
with all couplings encoded by two prepotentials. The remaining scalar fields
extend one of these manifolds into a QK or PQK manifold. These extensions
are unique, in the sense that the (P)QK manifold is obtained from an underlying 
S(P)K manifold by a c-map. Type-IIA and type-IIB differ in that 
in type-IIA the vector muliplets contain the (complexified or para-complexified) K\"ahler moduli, 
while in type-IIB the vector multiplets contain the complex structure moduli. 
The IIA hypermultiplets contain the complex structure moduli together with
the dilaton, the axion, and the R-R scalars. In type-IIA$_{(0,10)}$ and 
type-IIA$_{(1,9)}$  this leads to a QK
manifold, while for type-IIA$^*_{(1,9)}$ and for type-IIA$^{(2,8)}$ 
a sign flip for the R-R scalars leads to a $PQK_{SK}$ manifold. 
Type-IIB hypermultiplets contain the complexified K\"ahler moduli
for type-IIB/IIB$^*$, which then extends to a QK/PQK$_{SK}$ 
manifold, respectively. For type-IIB' the K\"ahler moduli space
is para-complexified, and extended to a $PQK_{SPK}$ manifold. 
See Table \ref{Tab:T-dualities} for a summary.

\subsection{Ten-dimensional Lagrangians}

In Section \ref{sec:type-II} we used the string frame parametrization of
\cite{Hull:1998vg} to display the various type-IIA Lagrangians.\footnote{Complete bosonic
string frame (pseudo-)Lagrangians for all type-II theories can be found in 
the appendix of  \cite{Dijkgraaf:2016lym}.}  In order 
to use the results of \cite{Bodner:1990zm} and \cite{Sabra:2015tsa}
on CY3 compactifications, we use the following Einstein frame
parametrization:
\begin{equation}
\label{eq:iias}
\begin{aligned}
		S_{IIA} = \int_{M_{10}} &\half \star R_{10} - \frac{9}{16} d \log \phi \wedge \star d \log \phi -
		\frac{\alpha_1}{4} \phi^{\frac{9}{4}} dV \wedge \star dV \\
		&- \frac{\alpha_2}{2} \phi^{-\frac{3}{2}} H_3 \wedge \star H_3 
		- \frac{\alpha_3}{2} \phi^{\frac{3}{4}} \left(F_4 + dV \wedge B_2 \right) \wedge \star \left(F_4 + dV \wedge B_2 \right) \\
		&- \frac{\sqrt{2}}{2} \left(F_4 + dV \wedge B_2 \right) \wedge F_4 \wedge B_2 - \frac{\sqrt{2}}{6} dV \wedge B_2 \wedge dV \wedge B_2 \wedge B_2 \;.
\end{aligned}
\end{equation}
Note that the dilaton has been redefined according to $\Phi \propto \log \phi$. Moreover we 
now use the same notations for form-fields as in \cite{Bodner:1990zm} and \cite{Sabra:2015tsa}:
$V$ is the RR one-form, while $F_4$ is the four-form field strength. The three parameters
$\alpha_i$, $i=1,2,3$ encode the various sign flips, see Table \ref{Tab:IIA_bis}.

\begin{table}
\begin{center}
\begin{tabular}{|l |r| r| r| } \hline
Type & $\alpha_1$ & $\alpha_2$ & $\alpha_3$ \\ \hline \hline
IIA$_{(1,9)}$ &1 &1 &1  \\ \hline
IIA$^*_{(1,9)}$ &$-1$ & 1  & $-1$\\ \hline
IIA$_{(0,10)}$ & $-1$ &$-1$  & 1 \\ \hline
IIA$_{(2,8)}$ &1  &$-1$ & $-1$ \\ \hline
\end{tabular}
\end{center}
\caption{Relative signs for kinetic terms in ten-dimensional type-IIA theories. 
A plus sign corresponds to a standard kinetic term in Lorentz signature,
as realized in IIA$_{(1,9)}$. \label{Tab:IIA_bis}}
\end{table}
As a quick check, note that this action is consistent with Table \ref{Tab:IIA}.
We also note that the three signs are not independent, since $\alpha_3 = \alpha_1 \alpha_2$.
This reflects that the three signs encode four independent theories, rather than six.
For type-IIA$_{(0,10)}$ and type-IIA$_{(1,9)}$ it was shown in \cite{Sabra:2015tsa}
that these Lagrangians arise from dimensional reduction of eleven-dimensional supergravity
with signature (1,10). For signature (1,9) one recovers the Lagrangian of 
\cite{Bodner:1990zm}.\footnote{As remarked
in \cite{Sabra:2015tsa} the second term in the third line is absent in \cite{Bodner:1990zm},
but present in \cite{Ferrara:1988ff}. It is straightforward to check that this term 
is generated by the field redefinition described explicitly in \cite{Sabra:2015tsa}.}
 The Lagrangians for type-IIA$^*_{(1,9)}$ and 
type-IIA$_{(2,8)}$ are obtained by a similar computation starting from the
Lagrangian for eleven-dimensional supergravity in signature (2,9) given 
in  \cite{Hull:1998vg}. Since these Lagrangians only differ by relative signs, and since in all cases
we reduce on the same manifold, we can use the results of 
\cite{Bodner:1990zm} and \cite{Sabra:2015tsa} for the individual terms, and
afterwards assemble them into four distinct four-dimensional Lagrangians.

\subsection{Reduction of the graviton-dilaton sector}

Since the Einstein-Hilbert and dilaton term are the same for all cases we can discuss 
them at once. 

Following  \cite{Bodner:1990zm} and  \cite{Sabra:2015tsa}  the reduction of 
\begin{equation*}
	S_{EH + \phi} = \int_{M_{10}} \half \star R_{10} - \frac{9}{16} d \log \phi \wedge \star \log \phi 
\end{equation*}
results in 
\begin{equation}
\label{eq:ehphi}
	S_{EH + \phi} = \int_{M_{4}} \half \star R_4 - \half G_{AB} (v) dv^A \wedge \star dv^B - \frac{1}{4} d \varphi \wedge \star d \varphi - g_{\alpha \bar{\beta}}(z,\bar{z}) dz^\alpha \wedge \star d \bar{z}^{\bar{\beta}} \;.
\end{equation}
Here $z^\alpha$ are the complex structure moduli with their special K\"ahler metric 
$g_{\alpha \beta}$ and $v^A$ are the real K\"ahler moduli with their special real
metric $G_{AB}(v)$, related to the previously introduced $M^A$ by the field redefinition
\begin{equation}
\label{eq:fieldredef}
	M^A = \sqrt{2} \phi^{-3/4} v^A \;.
\end{equation}
The four-dimensional dilaton $\varphi$ is related to the ten-dimensional dilaton $\phi$
by
\begin{equation*}
	\varphi = \log \left( 2 \V \phi^{-3} \right) .
\end{equation*}
Further details are given in \cite{Sabra:2015tsa}.

\subsection{Contribution of the $B$-field to the vector multiplet sector}

Next we turn to terms descending from the $B$-field kinetic term, which arise from 
taking the internal part of the $B$-field to be a harmonic two-form. 
Following \cite{Bodner:1990zm} and \cite{Sabra:2015tsa}
we say that terms which arise from harmonic two-forms belong to the $H^2$-cohomology sector. 
These are precisely the terms which contribute to the gravity plus vector multiplet sector
of the four-dimensional theory.
The ten-dimensional term takes the form
\begin{equation*}
	S_{H^2(B_2)} = \alpha_2\int_{M_{10}} - \half \phi^{-3/2} H_3 \wedge \star H_3 \bigg|_{H^2}
\end{equation*} 
where the sign depends on which theory we start with. We denote the projection of terms
onto the $H^2$-cohomology sector by $|_{H^2}$. 
Decomposing $\left. B_2 \right|_{H^2} = a^A V^A$ and 
 integrating over the CY3, we obtain
\begin{equation*}
	 S_{H^2(B_2)} =  \alpha_2 \int_{M_4} - \half G_{AB}(v) da^A \wedge \star da^B \; ,
\end{equation*}
where $a^A$ are four-dimensional scalar fields. We can combine this term
with one of the  terms obtained from the reduction of the Einstein-Hilbert term to obtain
\begin{equation*}
	   \int_{M_4} - \half G_{AB}(v) \left(dv^A \wedge \star dv^B + \alpha_2  da^A \wedge \star 
	   da^B \right) \;. 
\end{equation*} 
Making the field redefinition
\begin{equation}
\label{eq:fieldredef2}
	v^A = \frac{1}{2^{1/6}} y^A, \qquad a^A = - \frac{1}{2^{1/6}} x^A, \qquad \mathcal{K}_{ABC} = 
	c_{ABC} \;,
\end{equation}
we can rewrite this contribution as
\begin{equation*}
	 \int_{M_4} - \bar{g}_{AB}(y) \left(dx^A \wedge \star dx^B + \alpha_2 dy^A \wedge \star dy^B \right) \; 
\end{equation*}
where we have defined the new coupling matrix by
\begin{equation}
\label{little_g}
	\bar{g}_{AB} := \half \alpha_2 G_{AB} = - \frac{3}{2}  \alpha_2 \left( \frac{(cy)_{AB}}{(cyyy)} - \frac{3}{2} \frac{(cyy)_A (cyy)_B}{(cyyy)^2} \right) \;.
\end{equation}

\subsection{Contribution of R-R-kinetic terms to the vector multiplet sector}

Next we consider the contribution of the kinetic term of the R-R three-form, including 
its Chern-Simons like improvement term, to the $H^2$-sector, that is to 
the four-dimensional vector multiplets.
Depending on which theory we start with, the ten-dimensional term is
\begin{equation*}
	S_{H^2}(A_3) = \alpha_3  \int_{M_{10}} \frac{1} {2} \phi^{3/4} \left(F_4 + dV \wedge B_2 \right) \wedge \star \left(F_4 + dV \wedge B_2 \right) \bigg|_{H^2}.
\end{equation*}
Following \cite{Sabra:2015tsa} we  decompose $F_4$ and $B_2$ as
\begin{equation}
\label{eq:H2decomp}
	F_4 \big|_{H^2} = \F^A \wedge V^A, \qquad B_2 \big|_{H^2} = a^A V^A,
\end{equation}
where $\F^A$ are four-dimensional field strengths,
$\F^A = d\A^A$. Inserting this into the above action, we obtain an action ready to integrate over:
\begin{equation*}
	S_{H^2}(A_3) =\alpha_3  \int_{M_4} \frac{1}{2} \phi^{3/4} \left(\F^A + a^A dV \right) \wedge \star \left(\F^B + a^B dV \right) \int_{CY_3} V^A \wedge \star V^B \; .
\end{equation*}
Performing the integral we obtain 
\begin{equation*}
	S_{H^2}(A_3) = \alpha_3  \int_{M_4} \frac{\sqrt{2}}{3!} \K(v) G_{AB} (v) \left(\F^A + a^A dV \right) \wedge \star \left(\F^B + a^B dV \right) \bigg|_{H^2}  \;.
\end{equation*}

Similarly, the reduction of the kinetic term of the R-R one-form
\begin{equation*}
	S_V = \alpha_1  \int_{M_{10}} \frac{1}{4} \phi^{\frac{9}{4}} dV \wedge \star dV 
\end{equation*}
becomes 
\begin{equation*}
	S_V =  \alpha_1 \int_{M_4} \frac{ \sqrt{2}}{2 \cdot 3!} \K(v) \F^0 \wedge \star \F^0
\end{equation*}
after integration over the CY3, where we have set $\F^0 = dV$ and used the field redefinition (\ref{eq:fieldredef}).

%%%%

Using that $\alpha_3=\alpha_1 \alpha_2$, 
we can combine terms as 
\begin{equation*}
\begin{aligned}
	S_{H^2}(A_3) + S_V = & \alpha_1 \int_{M_4} \sqrt{2}  \left(\frac{1}{12} (\K vvv) + \frac{\alpha_2}{6} (\K vvv) G_{AB} a^A a^B \right) \F^0 \wedge \star \F^0\\
	&+ \frac{\sqrt{2} \alpha_2 }{3} (\K vvv) G_{AB} a^B \F^A \wedge \star \F^0 + \frac{\sqrt{2}\alpha_2 }{6} (\K vvv) G_{AB} \F^A \wedge \star \F^B \;,
\end{aligned}
\end{equation*}
where $(\K vvv) := \K_{ABC} v^A v^B v^C$. Rescaling the gauge fields
\begin{equation*}
	\F^A = \frac{1}{2^{1/6}} F^A
\end{equation*}
as well as the scalars, 
and using (\ref{little_g}), 
we can express the above as
\begin{equation*}
\begin{aligned}
	S_{H^2}(A_3) + S_V = & \alpha_1 \int_{M_4} \frac{1}{2} (c yyy) \left(\frac{1}{6} + \frac{2}{3} (g xx) \right) F^0 \wedge \star F^0\\
	&- \frac{2 }{3} (cyyy) (gx)_A F^A \wedge \star F^0 + \frac{1}{3} (c yyy) g_{AB} F^A \wedge \star F^B\;.
\end{aligned}
\end{equation*}

\subsection{Contribution of the 
topological terms to the vector multiplet sector}

The final contribution to the gravity plus vector multiplet sector
comes from the topological terms
\begin{equation*}
	S_{H^2(top)} = \int_{M_{10}} - \frac{\sqrt{2}}{2} \left(F_4 + dV \wedge B_2 \right) \wedge F_4 \wedge B_2 - \frac{\sqrt{2}}{6} dV \wedge B_2 \wedge dV \wedge B_2 \wedge B_2 \bigg|_{H^2} \;.
\end{equation*}
According to \cite{Sabra:2015tsa}, after reduction and field redefinitions this takes the form
\begin{equation*}
	S_{H^2(top)} = \int_{M_4} \frac{1}{6}\left[ 3(cx)_{AB} \F^A \wedge \F^B - 3(cxx)_A \F^A \wedge \F^0 + (cxxx) \F^0 \wedge \F^0 \right] .
\end{equation*}

\subsection{Final result for the gravity and vector multiplet sector}

Combining everything obtained so far gives us the bosonic Lagrangian for the gravity 
multiplet and the vector multiplets:
\begin{equation*}
\begin{aligned}
	S_{G + VM} = &\int_{M_4} \half \star R_4 - \bar{g}_{AB}(y) \left(dx^A \wedge \star dx^B + \alpha_2 dy^A \wedge \star dy^B \right) \\
&- \alpha_1 \bigg[\frac{1}{2} (c yyy) \left(\frac{1}{6} + \frac{2}{3} (g xx) \right) \F^0 \wedge \star \F^0\\
&- \frac{2}{3} (cyyy) (gx)_A \F^A \wedge \star \F^0 + \frac{1}{3} (c yyy) g_{AB} \F^A \wedge \star \F^B \bigg] \\
&+ \frac{1}{6}\left[ 3(cx)_{AB} \F^A \wedge \F^B - 3(cxx)_A \F^A \wedge \F^0 + (cxxx) \F^0 \wedge \F^0 \right] .
\end{aligned}
\end{equation*}
As shown in \cite{Cortes:2009cs}, one can introduce 
complex fields $z^A = x^A + i y^A$ if $\alpha_2=1$ and
para-complex fields $z^A = x^A + e y^A$ if $\alpha_2=-1$.
Then the second term in the first line becomes
a sigma model 
\begin{equation*}
\int_{M_4} - \bar{g}_{A\bar{B}}(z,\bar{z}) dz^A \wedge \star d\bar{z}^{\bar B}
\end{equation*}
with a target
space which is projective special K\"ahler for $\alpha_2=1$ and
projective special para-K\"ahler for $\alpha_2=-1$.
More generally, the results of \cite{Cortes:2009cs} imply 
that the full Lagrangian can be rewritten into the form
\begin{equation*}
\begin{aligned}
	S_{G + VM} = \int_{M_4} \half &\star R_4 - \bar{g}_{A\bar{B}} (z, \bar{z}) dz^A \wedge \star d \bar{z}^{\bar{B}} + \frac{\alpha_1}{4} \I_{\Sigma \Lambda} F^\Sigma \wedge \star F^\Lambda + \frac{1}{4} \cR_{\Sigma \Lambda} 
	F^\Sigma \wedge F^\Lambda \;,
\end{aligned}
\end{equation*}
where $\Sigma, \Lambda  = 0, 1, \ldots, n_V=h_{1,1}$, and 
where the vector coupling matrices $\mathcal{I}_{\Lambda \Sigma}$ and
$\mathcal{R}_{\Lambda \Sigma}$ can be expressed by a holomorphic or para-holomorphic
prepotential through the standard formulae of special geometry. This completes the 
derivation of the bosonic gravity plus vector multiplet Lagrangians for the four
type-IIA theories. Our result indeed matches \eqref{4d_VM_Lagrangian_IIA/IIA-star}
and \eqref{4d_VM_Lagrangian_para}.
For signature (1,3), where $\alpha_2=1$ and $\alpha_1=\alpha_3=
-\lambda = \pm 1$ we obtain a complex, (projective) SK scalar manifold, and the sign of
the Maxwell term distinguishes between type-IIA and type-IIA$^*$. For 
signatures (0,4) and (2,2), where $\alpha_2 = -1$, we obtain a para-complex, 
(projective) SPK manifold, and both signatures differ by the sign of the Maxwell
term, which is controlled by $\alpha_1 = - \alpha_3=-\lambda$. However, in these signatures
the supersymmetry algebra is unique and the sign can be changed by a 
field redefinition \cite{Cortes:2019mfa}.

\subsection{Contribution of the kinetic R-R-terms to the hypermultiplet sector}

We now turn to contributions from terms where the internal part is a harmonic 
3-form. So far we have discussed one such term, which arises from the 
reduction of the Einstein-Hilbert term. This results in the sigma model 
$- \tilde{G}_{\alpha \bar{\beta}}(z,\bar{z}) dz^\alpha \wedge \star d \bar{z}^{\bar{\beta}}$,
where $z^\alpha$ parametrize the deformations of the complex structure of the CY3,
with (projective) SK metric $ g_{\alpha \bar{\beta}}(z,\bar{z}) $. 
The R-R one-form does not contribute, but there are contributions for the 
R-R three-form and from the Kalb-Ramond field. 

We start with the contribution of the kinetic term of the four-form 
field strength $F_4 = d A_3$,
\begin{equation*}
	S_{H^3}(A_3) = \int_{M_10} \frac{-\alpha_3}{2} \phi^{3/4} F_4 \wedge \star F_4 \bigg|_{H^3}.
\end{equation*}
It has been shown in \cite{Bodner:1990zm}, \cite{Sabra:2015tsa} that
%%%
 \begin{equation}
\label{eq:acheck}
	F_4 |_{H^3} = d \check{A} = 2^{1/4} d \zeta^I \wedge \alpha_I 
	+ 2^{1/4} d \tilde{\zeta}_I \wedge \beta^I =
		 P^I \wedge \Phi_I + \bar{Q} \wedge \bar{\Omega} + \text{h.c.}	
\end{equation}
where the complex one-forms $P^I, \bar{Q} \in \Omega^1(M_4) \otimes \mathbb{C}$
can be expressed in terms of special geometry data associated with complex structure moduli as
\begin{equation}
\label{eq:PQ}
	P^I = i 2^{1/4} \left(d \tilde{\zeta}_J + \N_{JK} d\zeta^K \right) N^{IJ}, \qquad \bar{Q} = -i 2^{1/4} \frac{X^I}{(XN\bar{X})} \left(d \tilde{\zeta}_I + \N_{IJ} d\zeta^J\right)	 \;.
\end{equation}

While the four-dimensional hypermultiplet scalars $\zeta^I, \tilde{\zeta}_I$ are defined 
through the expansion of $d\check{A}$ in terms of $\alpha_I, \beta^I$, the expansion 
of $d\check{A}$ in terms of $\Phi_I, \Omega$ is used to carry out the integration over the
CY3:
 \begin{equation*}
	S_{H^3}(A_3) = \int_{M_4} -\alpha_3 \phi^{3/4} P^I \wedge \star \bar{P}^J \int_{CY_3} \Phi_I \wedge \star \bar{\Phi}_J + \int_{M_4} \alpha_3 \phi^{3/4} \bar{Q} \wedge \star Q \int_{CY_3} \bar{\Omega} \wedge \star \Omega \;.
\end{equation*}
Following \cite{Sabra:2015tsa} this can be evaluated and ultimately brought to the form
\begin{equation*}
	S_{H^3}(A_3) = \int_{M_4}  \frac{\alpha_3}{2} e^{-\varphi} \left[ \I_{IJ} d\zeta^I \wedge \star d\zeta^J + \I^{IJ} \left(d \tilde{\zeta}_I + \cR_{IK} d \zeta^K \right) \wedge \star \left(d \tilde{\zeta}_J + \cR_{JK} d \zeta^K \right) \right]  \;,
\end{equation*}
where $\varphi$ is the four-dimensional dilaton.

%%%%%%%%%%%%%%%%

\subsection{Contribution of topological terms and of the $B$-field to the hypermultiplet sector}

The topological contribution to the $H^3$-sector comes from
\begin{equation*}
	S_{H^3(\text{top})} = \int_{M_{10}} - \frac{\sqrt{2}}{2} F_4 \wedge F_4 \wedge B_2 \bigg|_{H^3}
\end{equation*}
Inserting in the expansions of these fields, we obtain
\begin{equation*}
	S_{H^3(\text{top})} = \int_{M_{4}} - \sqrt{2} \mathcal{B}_2 \wedge P^I \wedge \bar{P}^J \int_{CY_3} \Phi_I \wedge \bar{\Phi}_J + \int_{M_{4}} \sqrt{2} \mathcal{B}_2 \wedge \bar{Q} \wedge Q \int_{CY_3} \bar{\Omega} \wedge \Omega \;.
\end{equation*}
Following \cite{Sabra:2015tsa} this can be evaluated and ultimately be brought to the form
\begin{equation}
	S_{H^3(\text{top})} = -\int_{M_{4}} 2 \mathcal{B}_2 \wedge d\zeta^I \wedge d \tilde{\zeta}_I
	= \int_{M_{4}} 2 \Ham_3 \wedge \zeta^I d \tilde{\zeta}_I \;,
\end{equation}
where $\Ham_3 = d \mathcal{B}_2$ is the field strength of the four-dimensional
Kalb-Ramond field $\mathcal{B}_2$. 

To this we add the contribution from the reduction of the kinetic term of the 
ten-dimensional Kalb-Ramond field. (Here the internal part is a zero-form,
so this belongs to the `$H^0$ sector.')
\begin{equation*}
	S_{H^3(\text{top})} + S_{H^0(B_2)} = \int_{M_{4}} 2 \Ham_3 \wedge \zeta^I d \tilde{\zeta}_I - 
	\alpha_2 e^{2\varphi} \Ham_3 \wedge \star \Ham_3 \;. 
	\end{equation*}
Following \cite{Sabra:2015tsa} we dualize the four-dimensional Kalb-Ramond field
$\mathcal{B}_2$ into a scalar field $\tilde{\phi}$.

\begin{equation*}
\begin{aligned}
	S_{H^3(\text{top})} + S_{H^0(\tilde{\phi})} &= - \int_{M_{4}} e^{-2 \varphi} \left[d \tilde{\phi} + \half \left(\zeta^I d \tilde{\zeta}_I - \tilde{\zeta} d \zeta^I \right) \right] \wedge \star \left[d \tilde{\phi} + \half \left(\zeta^I d \tilde{\zeta}_I - \tilde{\zeta} d \zeta^I \right) \right]  \;.
\end{aligned}
\end{equation*}
Note that $\alpha_2$ has cancelled, because in signatures (0,10) and (2,8),
where $\alpha_2=-1$, the Hodge dualization generates an additional sign compared
to signature (1,9), where $\alpha_2=1$. 

\subsection{Final result for the hypermultiplet sector}

By collecting all terms contributing to the hypermultiplet sector we obtain
\begin{equation*}
\begin{aligned}
	S_{H} = \int_{M_4} &-  \tilde{G}_{\alpha \bar{\beta}} d z^\alpha \wedge \star d \bar{z}^{\bar{\beta}} - \frac{1}{4} d \varphi \wedge \star d \varphi \\
	&- e^{-2 \varphi} 
		\left[d \tilde{\phi} + \half \left(\zeta^I d \tilde{\zeta}_I - \tilde{\zeta}_I d\zeta^I \right) \right] \wedge \star \left[d \tilde{\phi} + \half \left(\zeta^I d \tilde{\zeta}_I - \tilde{\zeta}_I d\zeta^I \right) \right] \\
		&+ \frac{\alpha_3}{2} e^{- \varphi} \left[ \I_{IJ} d \zeta^I \wedge \star d\zeta^J + \I^{IJ} \left(d \tilde{\zeta}_I + \cR_{IK} d \zeta^K \right)
		\wedge \star \left(d \tilde{\zeta}_I + \cR_{IK} d \zeta^K \right) \right]
\end{aligned}
\end{equation*}
where $\alpha, \beta=1, \ldots, h_{2,1}=n_V-1$ and $I,J=1, \ldots, n_V = h_{2,1}+1$. 
This indeed agrees with \eqref{HM_IIA} upon identifying $\alpha_3=-\lambda$.
This completes the derivation of the four type-IIA hypermultiplet Lagrangians 
from dimensional reduction. For type-IIA$_{(1,9)}$ and type-IIA$_{(0,10)}$, where
$\alpha_3=-\lambda = 1$, 
the geometry is QK ($\alpha_3=1$), while for type-IIA$^*{(1,9)}$ and type-IIA${(2,8)}$
where $\alpha_3=-\lambda =-1$, the geometry is PQK. In both cases the distinguished
submanifold is the 
SK manifold provided by the complex structure moduli space.

\providecommand{\href}[2]{#2}\begingroup\raggedright\endgroup

\end{document}